\documentclass[aps,prl,reprint,superscriptaddress,floatfix,nofootinbib,longbibliography]{revtex4-2}
\usepackage{amsmath,amssymb,amsthm,bm,booktabs,graphicx,lmodern,mathrsfs}
\usepackage{mathtools,mleftright}
\usepackage[hidelinks]{hyperref}
\mleftright
\newcommand{\bbC}{\mathbb{C}}
\newcommand{\bbF}{\mathbb{F}}
\newcommand{\bbR}{\mathbb{R}}
\newcommand{\bbZ}{\mathbb{Z}}

\newcommand{\bbone}{\mathbf{1}}

\newcommand{\caA}{\mathcal{A}}
\newcommand{\caB}{\mathcal{B}}
\newcommand{\caC}{\mathcal{C}}

\newcommand{\caH}{\mathcal{H}}

\newcommand{\caM}{\mathcal{M}}
\newcommand{\caN}{\mathcal{N}}
\newcommand{\caO}{\mathcal{O}}

\newcommand{\caS}{\mathcal{S}}

\newcommand{\scrI}{\mathscr{I}}

\newcommand{\sfP}{\mathsf{P}}
\newcommand{\sfS}{\mathsf{S}}
\newcommand{\sfT}{\mathsf{T}}

\DeclareMathOperator{\Beta}{Beta}
\DeclareMathOperator{\Cov}{Cov}
\DeclareMathOperator{\diag}{diag}
\DeclareMathOperator{\E}{\mathbb{E}}
\DeclareMathOperator{\Exp}{Exp}
\renewcommand{\Re}{\operatorname{Re}}

\DeclareMathOperator{\tr}{tr}

\newcommand{\MUB}{\mathrm{MUB}}
\newcommand{\QR}{\mathrm{QR}}
\newcommand{\op}{\mathrm{op}}
\newcommand{\reg}{\mathrm{reg}}
\newcommand{\rmd}{\mathrm{d}}
\newcommand{\rme}{\mathrm{e}}
\newcommand{\rmi}{\mathrm{i}}
\newcommand{\stdState}{\mathrm{std}}

\newcommand{\lsp}{\mkern 2mu}

\providecommand{\bra}[1]{\langle#1\rvert}
\providecommand{\ket}[1]{\lvert#1\rangle}
\newcommand{\ketbra}[2]{\ket{#1}\bra{#2}}
\newcommand{\norm}[1]{\lVert#1\rVert}
\newcommand{\proj}[1]{\ketbra{#1}{#1}}

\newcommand{\eref}[1]{Eq.~\textup{(\ref{#1})}}
\newcommand{\Eref}[1]{Equation~\textup{(\ref{#1})}}
\newcommand{\eqsref}[2]{Eqs.~(\ref{#1}) and (\ref{#2})}

\newcommand{\eqssref}[3]{Eqs.~(\ref{#1}), (\ref{#2}), and (\ref{#3})}

\newcommand{\fref}[1]{Fig.~\ref{#1}}
\newcommand{\Fref}[1]{Figure~\ref{#1}}

\newcommand{\lref}[1]{Lemma~\ref{#1}}
\newcommand{\Lref}[1]{Lemma~\ref{#1}}

\newcommand{\pref}[1]{Proposition~\ref{#1}}

\newcommand{\psref}[2]{Propositions~\ref{#1} and \ref{#2}}

\newcommand{\sref}[1]{Sec.~\ref{#1}}

\newcommand{\tref}[1]{Table~\ref{#1}}
\newcommand{\Tref}[1]{Table~\ref{#1}}

\newcommand{\thref}[1]{Theorem~\ref{#1}}
\newcommand{\Thref}[1]{Theorem~\ref{#1}}
\newcommand{\thsref}[2]{Theorems~\ref{#1} and \ref{#2}}

\newtheorem{lemma}{Lemma}
\newtheorem{proposition}{Proposition}
\newtheorem{theorem}{Theorem}

\begin{document}
\title{\texorpdfstring{Near-Optimal Multiparameter Pure-State Metrology \mbox{with a Constant Number of Bases}}{Near-Optimal Multiparameter Pure-State Metrology with a Constant Number of Bases}}
\author{Bowen Xiao}
\thanks{These authors contributed equally to this work.}
\author{Zhiyao Yang}
\thanks{These authors contributed equally to this work.}
\affiliation{State Key Laboratory of Surface Physics, Department of Physics, and Center for Field Theory and Particle Physics, Fudan University, Shanghai 200433, China}
\affiliation{Institute for Nanoelectronic Devices and Quantum Computing, Fudan University, Shanghai 200433, China}
\affiliation{Shanghai Research Center for Quantum Sciences, Shanghai 201315, China}
\author{Huangjun Zhu}
\email{zhuhuangjun@fudan.edu.cn}
\affiliation{State Key Laboratory of Surface Physics, Department of Physics, and Center for Field Theory and Particle Physics, Fudan University, Shanghai 200433, China}
\affiliation{Institute for Nanoelectronic Devices and Quantum Computing, Fudan University, Shanghai 200433, China}
\affiliation{Shanghai Research Center for Quantum Sciences, Shanghai 201315, China}
\affiliation{Hefei National Laboratory, Hefei 230088, China}

\begin{abstract}
	Multiparameter pure-state metrology is limited by the incompatibility of measurements optimized for different parameters. We characterize uniform sensitivity by the Fisher floor, the smallest fraction of the quantum Fisher information (QFI) retained over all pure states and parameter directions. We prove that a constant number of fixed orthonormal bases suffices for a nearly optimal Fisher floor, independently of the Hilbert-space dimension $d$. With high probability, five independent Haar bases already achieve a positive, dimension-independent floor, and larger constant families achieve approximate universal Fisher symmetry: they retain nearly one half of the QFI in every direction, the largest uniform fraction permitted by measurement incompatibility. The required number of bases is optimal up to a logarithmic factor in inverse accuracy. In contrast, $d+1$ mutually unbiased bases, when they exist, can minimize mean Fisher anisotropy among families of $d+1$ bases yet have a Fisher floor of at most $1/(d+1)$, showing that optimal average balance does not ensure uniform sensitivity.
\end{abstract}
\maketitle

\emph{Introduction}---How many fixed measurement bases are needed to estimate all parameters of an unknown pure quantum state? Quantum state estimation and metrology both seek to extract information from limited measurement data~\cite{MassarPopescu1995,GiovannettiLloydMaccone2011,EisertEtAl2020}. In multiparameter estimation, a measurement optimized for one parameter can be less sensitive to others, and a measurement tailored to one state can lose sensitivity elsewhere~\cite{Paris2009,LiuEtAl2020,DemkowiczGoreckiGuta2020,Matsumoto2002,RagyJarzynaDemkowicz2016,*RagyJarzynaDemkowicz2019Erratum,PezzeEtAl2017}. We ask what precision a small collection of bases can support when chosen independently of the state.

The quantum Fisher information (QFI) quantifies the sensitivity encoded in the state, while the classical Fisher information quantifies how much of that sensitivity a measurement extracts~\cite{Helstrom1967,Wootters1981,BraunsteinCaves1994}. The classical Cram\'er--Rao bound relates the extracted information to local estimation precision. In dimension $d$, the Gill--Massar bound limits the total information in QFI units to $d-1$ per copy~\cite{GillMassar2000}. Since there are $2d-2$ real tangent directions, a measurement retains at most one half of the QFI per direction on average, and hence at most one half uniformly. A Fisher-symmetric measurement can attain this value in every direction at a prescribed state~\cite{LiFerrieGrossKalevCaves2016}. A continuous covariant measurement attains it at every state, while a finite-outcome single-copy measurement cannot do so exactly~\cite{Hayashi1998,ZhuHayashi2018}.

We quantify uniform sensitivity by the Fisher floor: the smallest fraction of QFI retained over all pure states and tangent directions. A positive floor ensures sensitivity in every direction, including in any smooth pure-state submodel. Prior work has established that a few bases can identify a pure state~\cite{HeinosaariMazzarellaWolf2013,CHST2015,GoyenecheEtAl2015}. Four generic bases distinguish all pure states~\cite{MondragonVoroninski2015}, and explicit four-basis constructions are also known~\cite{Jaming2014,CHST2015}. Five explicit bases can distinguish a pure state from any other state, including mixed states~\cite{CHKST2016}. Identification alone, however, does not quantify how much information such small collections retain as the dimension grows.

Randomized measurements are used in quantum tomography~\cite{GrossEtAl2010,AcharyaGuta2017} and in estimating properties by classical shadows~\cite{HuangKuengPreskill2020,ElbenEtAl2023}. A Haar-distributed basis gives half the QFI on average at any fixed pure state~\cite{AcharyaKypraiosGuta2016,LuSha2026}. A uniform guarantee for a finite draw must also control the least sensitive state and direction for that draw. An exact complex projective $3$-design gives a dimension-independent Fisher floor~\cite{ZhouChen2026,DuEtAl2026,YangXiaoZhu2026}, but requires at least $d^2(d+1)/2$ elements~\cite{RoyScott2007}, hence at least $d(d+1)/2$ bases if formed from orthonormal bases. Here we seek a uniform guarantee from a dimension-independent number of orthonormal bases.

\begin{table*}[!tbp]
	\caption{Fisher floors for equally weighted orthonormal-basis measurements in dimension $d\ge2$. Haar bases are independent. Probabilities refer to the choice of bases; the entries apply simultaneously to all pure states and parameter directions. The MUB row gives the exact floor for a complete family. The Haar rows give lower bounds on the floor; the constant $c_L$ increases with $L$ from $c_5>4.8\times10^{-10}$ toward $1/4$. The near-half row assumes $0<\varepsilon<1/2$.} \label{tab:settings-floor}
	\centering\setlength{\tabcolsep}{10pt}\renewcommand{\arraystretch}{1.15}
	\begin{tabular}{llll}
		\toprule
		Measurement family & Settings & Fisher floor & Probability or scope\\
		\midrule
		Complete MUBs & $d+1$ & $1/(d+1)$ & Exact, when the family exists\\
		Haar bases & $L\ge5$ & $c_L$ in \eref{eq:five-basis-bound} & At least $1-\rme^{-(d+18)/2}$\\
		Haar bases & $L\ge71$ & $1/10$ & At least $1-\rme^{-L(d+10)/60}$\\
		Haar bases & $L\ge271$ & $1/4$ & At least $1-\rme^{-L(d+10)/271}$\\
		Haar bases & $L\ge12\varepsilon^{-2}\ln(\rme/\varepsilon)$ & $1/2-\varepsilon$ & At least $1-\rme^{-L(d+16)\varepsilon^2/24}$\\
		\bottomrule
	\end{tabular}
\end{table*}

A standard way to combine complementary measurement settings is to use mutually unbiased bases (MUBs): a vector from one basis gives equally likely outcomes in every other basis. A complete family consists of $d+1$ MUBs and is known to exist in prime-power dimensions~\cite{WoottersFields1989}. We show that such a family can achieve optimal mean Fisher balance among families of this size, yet its Fisher floor is at most $1/(d+1)$. This contrast explains why good average performance does not settle the uniform estimation problem.

We prove that five independent Haar bases, drawn once and used with equal probability, retain a positive fraction of the QFI in every direction at every pure state. The certified fraction is independent of dimension, and the failure probability decays exponentially with dimension. With 71 and 271 bases, the certified floor is at least $1/10$ and $1/4$, respectively, with high probability in every dimension. To retain at least $1/2-\varepsilon$ of the QFI uniformly, a number of Haar bases of order $\varepsilon^{-2}\ln(1/\varepsilon)$ suffices as $\varepsilon$ tends to zero. Conversely, arbitrary basis choices and weights require order $\varepsilon^{-2}$ bases when the guarantee must hold in arbitrarily large dimensions. Thus a fixed set of bases can approach Fisher symmetry to prescribed accuracy, with a setting count independent of dimension and optimal up to a logarithmic factor. \Tref{tab:settings-floor} collects the explicit bounds; proofs are in the Supplemental Material (SM)~\cite{Supplemental}.

\emph{Fisher information and its floor}---Let $\caH$ be a complex Hilbert space of dimension $d\ge2$. Pure-state kets are normalized throughout. A finite positive operator-valued measure (POVM) $\caM=\{E_j\}_j$ with nonzero effects gives probabilities $p_j:=\bra\psi E_j\ket\psi$ at $\ket\psi\in\caH$. A state is \emph{regular} for $\caM$ if every $p_j$ is positive, and \emph{singular} otherwise (zero POVM elements are omited). All uniform statements below include both classes. Choose a smooth local parametrization of pure states with normalized ket representative $\ket{\psi(\bm\theta)}$ and $2d-2$ real coordinates $\bm\theta=(\theta_1,\ldots,\theta_{2d-2})$. With $\partial_\alpha:=\partial/\partial\theta_\alpha$, the classical and quantum Fisher information matrices per copy are
\begin{equation} \label{eq:fisher-matrices}
	\begin{aligned}
		I_{\alpha\beta}&:=\sum_{j:p_j>0}\frac{\partial_\alpha p_j\,\partial_\beta p_j}{p_j},\\
		J_{\alpha\beta}&:=4\Re[\langle\partial_\alpha\psi|\partial_\beta\psi\rangle-\langle\partial_\alpha\psi|\psi\rangle\langle\psi|\partial_\beta\psi\rangle].
	\end{aligned}
\end{equation}
Write $I_{\caM,\psi}$ and $J_\psi$ for these matrices at $\ket\psi$. The ordinary-score convention in \eref{eq:fisher-matrices} omits zero-probability terms.

To compare information about different parameters, express the classical Fisher information in units set by the QFI. Define the normalized Fisher information matrix and the Fisher floor by
\begin{equation} \label{eq:fisher-floor}
	G_\psi:=J_\psi^{-1/2}I_{\caM,\psi}J_\psi^{-1/2},\quad \eta(\caM):=\inf_{\ket\psi\in\caH}\lambda_{\min}(G_\psi).
\end{equation}
The eigenvalues of this real symmetric matrix, called \emph{Fisher eigenvalues}~\cite{Zhu2015,Zhu2023}, are independent of the choice of local coordinates. Along a generalized eigenvector of $I_{\caM,\psi}$ relative to $J_\psi$, the corresponding eigenvalue is the retained fraction of QFI. Zero gives no first-order information, while one retains the full directional QFI. The floor is the smallest fraction over all pure states and parameter directions. The SM also represents $G_\psi$ as a real-linear operator on the pure-state tangent space, with the same spectrum. The Gill--Massar bound $\tr\left(J_\psi^{-1}I_{\caM,\psi}\right)\le d-1$ implies $\eta\le1/2$~\cite{GillMassar2000}.

A floor $\eta(\caM)\ge c>0$ is equivalent to $I_{\caM,\psi}\succeq cJ_\psi$ at every pure state. This has a direct statistical meaning. For a locally unbiased estimator based on $N$ independent copies, the classical Cram\'er--Rao bound is $\Cov(\widehat{\bm\theta})\succeq I_{\caM,\psi}^{-1}/N$. The floor bounds this inverse-information benchmark above by $c^{-1}J_\psi^{-1}/N$, that is, $1/(2c)$ times the Fisher-symmetric benchmark $2J_\psi^{-1}/N$. At a regular state, the inverse-information benchmark $I_{\caM,\psi}^{-1}/N$ is attainable to leading order under the usual regularity assumptions~\cite{Hayashi1998}.

At a regular state, a single basis has $d-1$ unit Fisher eigenvalues and $d-1$ zero eigenvalues: it extracts the full QFI in some directions and none in the others. The purpose of using several bases is to cover all directions without knowing the state in advance.

For specified bases $\caB_\ell=\{\ket{\varphi_{\ell k}}\}_{k=0}^{d-1}$, $\ell=0,\ldots,L-1$, let $q_\ell\ge0$ denote the probability of selecting $\caB_\ell$, where $\sum_{\ell=0}^{L-1}q_\ell=1$. Their mixture $\caM(q)$ has effects $E_{\ell k}:=q_\ell\proj{\varphi_{\ell k}}$. Both the bases and their probabilities are fixed independently of the unknown state. Each basis counts as one setting, and with positive probabilities the mixture has $Ld$ labeled outcomes. All logarithms are natural.

\emph{Average balance and uniform sensitivity}---A measurement can distribute information evenly on average yet leave particular states poorly resolved. To distinguish these properties, we consider \emph{mean Fisher anisotropy}, the state average of the spectral deviation from Fisher symmetry:
\begin{equation} \label{eq:mean-spec-deviation}
	\wp(\caM):=\frac{1}{2(d-1)}\int\tr\left[\left(G_\psi-\frac{\bbone}{2}\right)^2\right]\,\rmd\psi.
\end{equation}
Here $\rmd\psi$ is normalized Haar measure. At each regular state of a rank-one POVM, the $2d-2$ Fisher eigenvalues have arithmetic mean $1/2$, since $\tr G_\psi=d-1$. Thus $\wp$ is the Haar-state mean of the variance of the Fisher eigenvalues. A single basis has $\eta=0$ and $\wp=1/4$.

MUBs are natural candidates for balancing information: vectors $\ket\varphi$ and $\ket\phi$ from distinct bases have $|\langle\varphi|\phi\rangle|^2=1/d$, so measuring a vector of one basis in another gives uniform outcomes. The following result shows that they minimize mean anisotropy for a fixed number of bases.

\begin{proposition}[Mean Fisher anisotropy] \label{prop:mean-anisotropy}
	For the equal mixture $\caM$ of $L\ge1$ orthonormal bases of $\caH$,
	\begin{equation} \label{eq:average-deviation}
		\wp(\caM)\ge\frac{1}{4}\left[1-\frac{(L-1)d}{L}\ln\frac{d}{d-1}\right].
	\end{equation}
	For $L=1$, $\wp=1/4$. For $L\ge2$, equality holds precisely for pairwise mutually unbiased bases.
\end{proposition}

The weighted extension in SM \sref{supp:average-mub} shows that unequal probabilities only increase this lower bound, so equally weighted MUBs, when they exist, minimize $\wp$ among all mixtures of the same number of bases. For complete MUBs, equality gives $\wp=1/(8d)+\caO(d^{-2})$.

Despite this optimal average balance, MUBs can retain very little information in a particular parameter direction. Their Fisher floor is determined exactly by the basis probabilities.

\begin{proposition}[Fisher floor of complete MUBs] \label{prop:mub-floor}
	Let $\caB_0,\ldots,\caB_d$ be a complete MUB family on $\caH$ and let $\caM_{\MUB}(q)$ be its mixture with basis probabilities $q=(q_0,\ldots,q_d)$. Then
	\begin{equation} \label{eq:mub-floor}
		\eta(\caM_{\MUB}(q))=\min_\ell q_\ell.
	\end{equation}
	In particular, equal probabilities give $1/(d+1)$, and every proper subset of the same complete family has zero floor.
\end{proposition}
Uniform weights maximize the floor within this MUB family. Even at this maximum, the floor tends to zero with $d$: averaging the squared deviations over directions and states can conceal a small information fraction at a particular state and direction. The SM proves both propositions in \sref{supp:fisher-moments}, and End Matter identifies the MUB tangent directions associated with the small eigenvalue.

\begin{figure}[!tb]
	\centering
	\includegraphics[width=0.95\columnwidth]{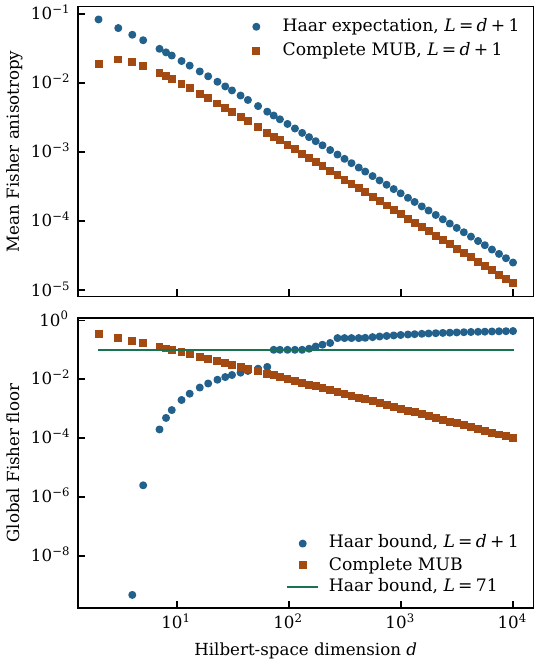}
	\caption{Mean Fisher anisotropy (top) and Fisher floor (bottom) at selected prime-power dimensions. MUB values are exact. For $L=d+1$ Haar bases, we plot the expected anisotropy and the largest evaluated floor lower bound from Theorems~\ref{thm:five-bases}--\ref{thm:near-half}; for each $d$, the theorem is selected before the bases are drawn. The horizontal line at $1/10$ is the lower bound for $L=71$ Haar bases from \thref{thm:large-floors}. All Haar floor bounds hold with probability at least $99.99\%$. Matched-count Haar floor points are omitted at $d=2,3$, where the theorems give no positive bound.} \label{fig:anisotropy-floor}
\end{figure}

Haar bases show the opposite trade-off. We write $\caM_{d,L}$ for the equal mixture of $L$ independent Haar bases, drawn once and then fixed. Probabilities refer to this initial draw, and each floor bound below holds simultaneously over all states and tangent directions on the corresponding event. On average, these measurements are less balanced than MUBs:
\begin{equation} \label{eq:haar-average-deviation}\E\wp(\caM_{d,L})=\frac{1}{4L},\end{equation}
as shown in SM \sref{supp:haar-anisotropy}. When both measurements use $L=d+1$ bases, the MUB anisotropy is asymptotically half the Haar expectation, as shown in the upper panel of \fref{fig:anisotropy-floor}. Yet, as we show next, five or more Haar bases give a floor bounded away from zero with high probability in every dimension.

\emph{Five bases suffice}---Can a fixed number of bases retain a positive fraction of the QFI at every pure state in every dimension $d\ge2$, with a bound independent of $d$? Our first theorem gives an affirmative answer with five independent Haar bases.
\begin{theorem}[Five-basis sufficiency] \label{thm:five-bases}
	For integers $d\ge2$ and $L\ge5$, the random measurement $\caM_{d,L}$ satisfies
	\begin{equation} \label{eq:five-basis-bound}
		\begin{gathered}
			\Pr\{\eta(\caM_{d,L})\ge c_L\}\ge1-\rme^{-(d+18)/2},\\
			c_L:=\frac{1}{4}\exp\left(-\frac{6L^{3/4}}{L-4}\right)>4.8\times10^{-10}.
		\end{gathered}
	\end{equation}
	The constant $c_L$ increases with $L$ and tends to $1/4$.
\end{theorem}

The failure bound is below $5\times10^{-5}$ in every dimension; SM \sref{supp:five-bases} also gives the stronger five-basis floor $10^{-8}$. These conservative certificates establish uniform positivity. Numerical minimization suggests that the actual floors are much larger: for five Haar bases, the numerically estimated floor, averaged over random draws, exceeds $7\times10^{-4}$ in every dimension $d\le100$.

\emph{Larger Fisher floors}---Increasing the number of settings gives substantial information fractions, still without dependence on dimension.
\begin{theorem}[Floors $1/10$ and $1/4$] \label{thm:large-floors}
	For integers $d\ge2$ and $L$, the random measurement $\caM_{d,L}$ satisfies
	\begin{equation} \label{eq:large-floor-bounds}
		\begin{aligned}
			&\!\!\Pr\left\{\eta(\caM_{d,L})\ge\frac{1}{10}\right\}\ge1-\rme^{-L(d+10)/60},\quad L\ge71,\\
			&\!\!\Pr\left\{\eta(\caM_{d,L})\ge\frac{1}{4}\right\}\ge1-\rme^{-L(d+10)/271},\quad L\ge271. 
		\end{aligned}
	\end{equation}
\end{theorem}

The floors $1/10$ and $1/4$ keep the inverse-information benchmark within factors $5$ and $2$, respectively, of the Fisher-symmetric value in every parameter direction. At the threshold counts, both failure probabilities are below $10^{-5}$ for every $d\ge2$. Unlike the floor $1/(d+1)$ of complete MUBs, these floors do not decay with dimension, as illustrated in the lower panel of \fref{fig:anisotropy-floor}. The quarter floor also confines every Fisher eigenvalue to $[1/4,3/4]$, as the following complementarity argument shows.

\emph{Fisher symmetry}---In the full pure-state model, Fisher symmetry means $I_{\caM,\psi}=J_\psi/2$: information is distributed equally among directions and saturates the Gill--Massar bound. Finite measurements can attain this balance at a prescribed pure state~\cite{LiFerrieGrossKalevCaves2016}. Exact universal Fisher symmetry for a single-copy measurement requires infinitely many outcomes~\cite{ZhuHayashi2018}. We ask how closely a fixed measurement can approach this balance everywhere.

For $0\le\varepsilon<1/2$, we say that $\caM$ has \emph{$\varepsilon$-approximate universal Fisher symmetry} if
\begin{equation} \label{eq:approx-symmetry}
	\left(\frac{1}{2}-\varepsilon\right)J_\psi\preceq I_{\caM,\psi}\preceq\left(\frac{1}{2}+\varepsilon\right)J_\psi\quad\forall\,\ket\psi\in\caH.
\end{equation}
Thus every Fisher eigenvalue differs from $1/2$ by at most $\varepsilon$. To see why a lower bound alone ensures this balance, consider a unit tangent ket $\ket\nu$ orthogonal to $\ket\psi$. For any finite POVM, the variations along $\ket\nu$ and $\rmi\ket\nu$ retain information fractions whose sum is at most one~\cite{YangXiaoZhu2026}. Since both fractions are at least $\eta(\caM)$, neither exceeds $1-\eta(\caM)$. Hence \eref{eq:approx-symmetry} is equivalent to $\eta(\caM)\ge1/2-\varepsilon$; SM \sref{supp:symmetry-proof} gives the short derivation.

\begin{theorem}[Approximate universal Fisher symmetry] \label{thm:near-half}
	Let $0<\varepsilon<1/2$. For integers $d\ge2$ and $L\ge12\varepsilon^{-2}\ln(\rme/\varepsilon)$, the random measurement $\caM_{d,L}$ satisfies
	\begin{equation} \label{eq:near-half}
		\Pr\left\{\eta(\caM_{d,L})\ge\frac{1}{2}-\varepsilon\right\}\ge1-\rme^{-L(d+16)\varepsilon^2/24}.
	\end{equation}
	On this event, \eref{eq:approx-symmetry} holds at all pure states.
\end{theorem}

At the stated count, the failure bound is below $10^{-6}$ for every $d\ge2$. The same measurement then gives nearly balanced information at every pure state. Its inverse-information benchmark is at most $1/(1-2\varepsilon)$ times the Fisher-symmetric value. Thus the precision penalty can be made arbitrarily small by choosing the accuracy $\varepsilon$, with no dimension dependence in the required basis count. For fixed accuracy, the measurement has only $Ld=\caO(d)$ outcomes, compared with the $2d-2$ real state parameters.

The gain from additional bases also holds in expectation. Define $\overline{\eta}_{\mathrm H}(d,L):=\E\eta(\caM_{d,L})$.

\begin{proposition}[Expected Fisher floor] \label{prop:mean-floor}
	For integers $d\ge2$ and $L\ge1$, the expected floor is nondecreasing in $L$ and satisfies
	\begin{equation} \label{eq:mean-floor}
		\overline{\eta}_{\mathrm H}(d,L)\ge\frac{1}{2}-\sqrt{\frac{6\ln(\rme L)}{L}}.
	\end{equation}
\end{proposition}

The expected floor therefore converges monotonically to $1/2$ uniformly in dimension. For moderate counts, \thsref{thm:five-bases}{thm:large-floors} give stronger expectation bounds. SM \sref{supp:average-floors} proves these bounds and \pref{prop:mean-floor}.

\begin{figure}[!tb]
	\centering
	\includegraphics[width=0.95\columnwidth]{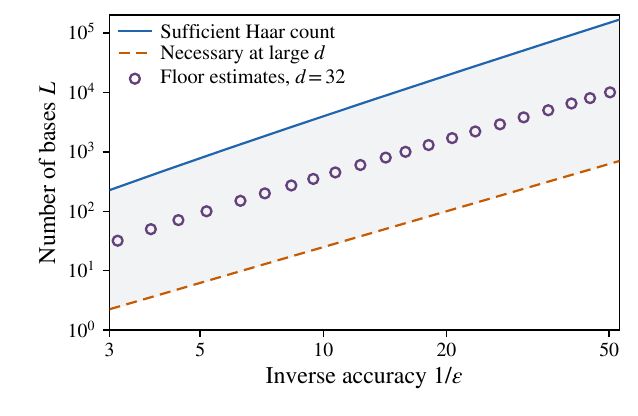}
	\caption{Number of measurement bases $L$ versus inverse accuracy $1/\varepsilon$ for a target Fisher floor of at least $1/2-\varepsilon$. The solid curve shows the sufficient number $12\varepsilon^{-2}\ln(\rme/\varepsilon)$ of Haar bases from \thref{thm:near-half}, at or above which the failure probability is below $10^{-6}$ in every dimension $d\ge2$. The dashed curve shows the necessary number $1/(4\varepsilon^2)$ from \pref{prop:basis-ceiling} for targets achieved in arbitrarily large dimensions; the shaded area marks the gap. The curves are not rounded to integers. Open circles show numerical results at $d=32$, plotted at $1/\varepsilon=1/(1/2-\bar{\eta}_L)$, where $\bar{\eta}_L$ is the mean floor estimate over 100 independent draws of $L$ Haar bases, each obtained from 800 search starts. Each estimate is an upper estimate of the floor of its draw.} \label{fig:accuracy-count}
\end{figure}

\emph{How many bases are necessary?}---The sufficient counts above leave open how many bases are unavoidable. We obtain a lower bound that applies to deterministic and random families alike, with arbitrary measurement probabilities.
\begin{proposition}[Limits imposed by the basis count] \label{prop:basis-ceiling}
	Let $\caM$ be a mixture of $L\ge1$ orthonormal bases of $\caH$, with arbitrary probabilities. Then $\eta(\caM)=0$ for $L\le2$, and
	\begin{equation} \label{eq:finite-ceiling}
		\eta(\caM)\le\frac{1}{2}-\frac{1}{2}\max\left\{\frac{1}{L},\frac{1}{\sqrt{L}\lsp }\left[1-\frac{L-1}{2(d-1)}\right]\right\}.
	\end{equation}
\end{proposition}
The $1/L$ term excludes exact universal Fisher symmetry for any finite basis count. At fixed $L$, the second term gives a gap of order $L^{-1/2}$ in the large-dimension limit. Let $\eta^*(d,L)$ be the supremum of $\eta(\caM)$ over all mixtures $\caM$ of $L$ orthonormal bases in dimension $d$, with arbitrary selection probabilities. For each fixed $L$,
\begin{equation} \label{eq:fixed-settings-ceiling}
	\limsup_{d\rightarrow\infty}\eta^*(d,L)\le\frac{1}{2}-\frac{1}{2\sqrt{L}\lsp }.
\end{equation}
Consequently, a floor $1/2-\varepsilon$ in arbitrarily large dimensions requires $L\ge1/(4\varepsilon^2)$. The sufficient count in \thref{thm:near-half} differs only by a logarithmic factor in $1/\varepsilon$. SM \sref{supp:basis-ceilings} gives the proof. End Matter develops the consequences for unequal probabilities and fixed dimension, and a separate probability bound for Haar families. The finite-dimension correction matters: the three qubit MUBs have floor $1/3$, exceeding the limiting ceiling $1/2-1/\bigl(2\sqrt{3}\lsp \bigr)$.

\Fref{fig:accuracy-count} compares the two counts as functions of the target accuracy. The sufficient count holds in every dimension, whereas the necessary count applies when the target must be achieved in arbitrarily large dimensions. Both grow quadratically in inverse accuracy; the sufficient count carries an additional logarithmic factor and a larger constant. At $d=32$, the numerical estimates suggest that substantially fewer bases may suffice to reach a given accuracy.

\emph{Summary}---We have shown that five Haar bases suffice for a dimension-independent positive Fisher floor. With 71 and 271 bases, the certified floor rises to $1/10$ and $1/4$, respectively, and a basis count independent of dimension brings it arbitrarily close to the optimal uniform value $1/2$. Our converse fixes the quadratic dependence of the basis count on inverse accuracy up to a logarithmic factor. Complete equal-weighted MUBs illustrate the distinction between average balance and uniform sensitivity: they minimize mean Fisher anisotropy among families of their size, yet their floor decreases as $1/(d+1)$. In quantum state estimation, infidelity weights tangent directions by the QFI to leading order. Hence, at each regular state, $\varepsilon$-approximate universal Fisher symmetry bounds the leading mean-infidelity coefficient by $(d-1)/(1-2\varepsilon)$ for asymptotically efficient estimators (SM \sref{supp:directional-proof}), compared with the optimal worst-case coefficient $d-1$ of collective measurements on all copies~\cite{MassarPopescu1995,Hayashi1998}. A constant number of fixed bases can thus nearly match even the most powerful multi-copy collective measurements.

A positive floor also gives a uniform finite-sample guarantee. For a fixed draw with floor at least $c>0$, the likelihood reconstruction bound of Ref.~\cite{YangXiaoZhu2026} gives worst-case mean infidelity at most $2(Ld-1)/(c^2N)$ for pure-state maximum-likelihood estimation from $N$ independent copies. This is $\caO(d/N)$ for fixed $L$ and $c$. It remains open whether four bases suffice for a dimension-independent positive floor and whether the logarithmic gap in the accuracy dependence can be closed. Constructing explicit bases with comparable guarantees is another open problem.

\emph{Note added}---After obtaining our main results, we learned of the preprint by W.~Li and J.~Liu (\mbox{\href{https://arxiv.org/abs/2611.00599}{arXiv:2611.00599}}). They also derive the floor and mean anisotropy of equally weighted complete MUBs and prove a positive floor for five Haar bases in large dimensions. Our five-basis floor applies for every $d\ge2$, and even the constant $c_5>4.8\times10^{-10}$ in \thref{thm:five-bases} exceeds their explicit lower bound by more than 20 orders of magnitude. We also give a dimension-independent count for approximate universal Fisher symmetry, optimal up to a logarithmic factor.

\emph{Acknowledgments}---We thank Luyu Lu, Datong Chen, Jianxiang Ma, and Xinyang Shu for inspiring discussions. This work is supported by National Natural Science Foundation of China (Grant No.\ 92576101),  Shanghai Science and Technology Innovation Action Plan (Grant No.\ 24LZ1400200), Quantum Science and Technology-National Science and Technology Major Project (Grant No.\ 2024ZD0300101), and National Key Research and Development Program of China (Grant No.\ 2022YFA1404204).

\makeatletter\let\SavedAddContentsLine\addcontentsline\let\addcontentsline\@gobblethree\makeatother
\begingroup 
\bibliography{Ref}
\endgroup\let\addcontentsline\SavedAddContentsLine
\par\bigskip
\begin{center}\textbf{End Matter}\end{center}
We first derive the Fisher spectrum of the MUB example, then discuss mixing and expected-floor monotonicity, limits on the basis count for arbitrary probabilities, and a lower-tail bound for Haar bases. A computable lower bound on the floor completes the discussion.

\emph{Fisher spectrum of MUBs in odd prime dimension}---This spectrum shows how nearly all Fisher eigenvalues can lie close to $1/2$ while one is small, here at a superposition of two computational-basis states.
\begin{proposition}[Fisher spectrum for odd-prime MUBs] \label{prop:mub-spectrum}
	Let $d$ be an odd prime and put $\omega=\rme^{2\pi\rmi/d}$. Define the bases $\caB_\ell=\{\ket{\varphi_{\ell k}}\}_{k=0}^{d-1}$, $\ell=0,\ldots,d-1$, by
	\begin{equation*}
		\ket{\varphi_{\ell k}}:=\frac{1}{\sqrt{d}\lsp}\sum_{j=0}^{d-1}\omega^{\ell j(j-1)+kj}\ket j.
	\end{equation*}
	Let $\caM_{\MUB}$ be the POVM uniformly mixing these bases and the computational basis $\caB_d:=\{\ket j\}_{j=0}^{d-1}$. At $\ket{\psi_+}:=(\ket0+\ket1)/\sqrt{2}$, its normalized Fisher spectrum is
	\begin{equation} \label{eq:prime-mub-spectrum}
		\frac{1}{d+1},\quad \frac{d}{2(d+1)},\quad \frac{d}{d+1},
	\end{equation}
	with multiplicities $1$, $2d-4$, and $1$, respectively.
\end{proposition}
The smallest eigenvalue attains the floor in \pref{prop:mub-floor}. The corresponding tangent ket $\ket{\nu_0}:=(\ket0-\ket1)/\sqrt{2}$ changes the relative populations of the two occupied computational-basis states. The conjugate phase direction $\rmi\ket{\nu_0}$ has the largest eigenvalue. The remaining eigenspace consists of tangent kets supported on $\ket2,\ldots,\ket{d-1}$. Each such complex amplitude supplies two real directions with eigenvalue $d/[2(d+1)]$. The infimum over regular states is also $1/(d+1)$ by \eref{eq:regular-floor}. The spectrum is derived in SM \sref{supp:mub-spectrum}.

\emph{Mixing and monotonicity of the expected floor}---Recording the measurement label together with the outcome makes the Fisher information of a mixture the weighted average of those of its components. Taking the infimum gives concavity of the floor, and averaging over one-POVM deletions gives monotonicity of the expected floor.
\begin{proposition}[Mixing measurements] \label{prop:mean-monotone}
	Let $\caN_j:=\{E_{ji}\}_i$, $j=1,\ldots,m$, be finite POVMs on $\caH$. For probabilities $q_j\ge0$ with $\sum_jq_j=1$, their mixture $\caM(q):=\{q_jE_{ji}\}_{j,i}$ satisfies
	\begin{equation} \label{eq:floor-concavity}
		\eta(\caM(q))\ge\sum_{j=1}^mq_j\eta(\caN_j).
	\end{equation}
	Suppose further that $m\ge2$ and that $\caN_1,\ldots,\caN_m$ are random POVMs whose joint distribution is invariant under permutations. Let $\caM_k$ denote the equal mixture of $\caN_1,\ldots,\caN_k$. Then
	\begin{equation*}
		\E\eta(\caM_m)\ge\E\eta(\caM_{m-1}).
	\end{equation*}
\end{proposition}
For a given family of bases, $\caM(q)$ is the mixture of the main text. Independent identically distributed POVMs, such as those based on independent Haar bases, satisfy the exchangeability condition. For the proof, $\caM_m$ has the same directional Fisher response as the equal mixture of its $m$ one-POVM deletions, so \eref{eq:floor-concavity} bounds its floor below by their average floor; exchangeability gives every deletion the expected floor of $\caM_{m-1}$. Monotonicity holds only in expectation: for a particular family, adding a basis can reduce the floor. For instance, repeating one of the three equally weighted qubit MUBs reduces the floor from $1/3$ to $1/4$ by \pref{prop:mub-floor}.

\emph{Fisher floor of $L$ bases}---Every mixture $\caM(q)$ of $L$ bases satisfies
\begin{equation} \label{eq:weighted-ceiling}
	\eta(\caM(q))\le\frac{1-\max_\ell q_\ell}{2}.
\end{equation}
At a vector of a most heavily weighted basis, that basis supplies no first-order information. The remaining probability $1-\max_\ell q_\ell$ must cover all tangent directions. For $0<\varepsilon<1/2$, achieving $\eta(\caM(q))\ge1/2-\varepsilon$ requires $\max_\ell q_\ell\le2\varepsilon$ and
\begin{equation} \label{eq:simple-basis-count}
	L\ge\max\left\{\frac{1}{2\varepsilon},\frac{2d-1}{1+8(d-1)\varepsilon^2}\right\}.
\end{equation}
The second count approaches $1/(4\varepsilon^2)$ when $d\varepsilon^2$ is large, while retaining the correction needed at small dimension. SM \sref{supp:basis-ceilings} proves these bounds and a sharper weighted ceiling.

These are necessary conditions for any family, regardless of how its bases are chosen. Nearly equal probabilities avoid the loss caused by concentrating the measurement on one basis, but they do not ensure a large floor: the equally weighted MUB example already shows that the relative geometry of the bases also matters.

\emph{Lower-tail probability of the Haar Fisher floor}---For Haar families, the following bound limits the confidence of a near-half floor at fixed dimension, complementing the deterministic large-dimension ceiling in \pref{prop:basis-ceiling}.
\begin{proposition}[Lower tail of the Haar Fisher floor] \label{prop:haar-converse}
	Let $d\ge2$ and $L\ge1$ be integers. For $0<\varepsilon<1/\bigl(2\sqrt{Ld}\lsp \bigr)$,
	\begin{equation} \label{eq:anti-concentr}
		\Pr\left\{\eta(\caM_{d,L})<\frac{1}{2}-\varepsilon\right\}\ge\frac{(1-4Ld\varepsilon^2)^2}{3}.
	\end{equation}
	For any target accuracy $0<\varepsilon<1/2$, if a fixed count $L$ gives a floor of at least $1/2-\varepsilon$ with probability at least $23/24$ in each dimension $d\ge2$, then $L>7/(80\varepsilon^2)$.
\end{proposition}
Already for qubits, success probability $23/24$ requires a quadratic count in inverse accuracy. SM \sref{supp:haar-converse} proves both statements.

This is a lower bound on the probability of falling below the target floor, complementing the upper failure bounds in our sufficiency theorems. Its proof needs only one fixed state and two phase-related tangent directions: a sufficiently large fluctuation in either direction already forces the uniform floor below the target. No minimization over all states is needed for this converse.

\emph{A computable lower bound on the Fisher floor}---For a specified measurement, a frame operator from informationally complete tomography~\cite{Scott2006,RoyScott2007} certifies a Fisher lower bound through a matrix eigenvalue. Its limitation is the number of bases required.
Let $\caM$ be the equal mixture of $L\ge1$ orthonormal bases $\caB_\ell=\{\ket{\varphi_{\ell k}}\}_{k=0}^{d-1}$ of $\caH$, with $\ell=0,\ldots,L-1$. Put $P_{\ell k}:=\proj{\varphi_{\ell k}}$. For Hermitian $O$, define
\begin{equation} \label{eq:frame-map}
	\sfS_{\caM}(O):=\frac{1}{L}\sum_{\ell=0}^{L-1}\sum_{k=0}^{d-1}\tr(P_{\ell k}O)P_{\ell k}.
\end{equation}
On the real vector space of traceless Hermitian operators, use the Hilbert--Schmidt inner product $\langle O,Y\rangle=\tr(OY)$. The map preserves this space and is self-adjoint. Its smallest eigenvalue on this space is
\begin{equation} \label{eq:frame-minimum}
	s_{\caM}:=\min_{\substack{O=O^\dagger,\ \tr O=0\\\tr(O^2)=1}}\tr[O\sfS_{\caM}(O)].
\end{equation}
\begin{proposition}[A frame certificate] \label{prop:frame-cert}
	For the equal mixture $\caM$ above, $\eta(\caM)\ge s_{\caM}$, with $s_{\caM}\le1/(d+1)$ and $s_{\caM}=0$ if $L<d+1$.
\end{proposition}

The test requires the smallest eigenvalue of a real matrix of size $d^2-1$, and gives only the trivial lower bound when $L<d+1$. Even above this count, $s_{\caM}\le1/(d+1)$ prevents a dimension-independent certificate. The test probes all traceless Hermitian operators, whereas pure-state information involves only rank-two tangents. Exploiting this smaller set, as in the five-basis proof, is essential for a certificate of constant strength in large dimension.

The strength of this test is its computability. Evaluating the floor itself, or the rank-two certificate in SM \sref{supp:deterministic-certs}, requires minimizing over pure states and tangent directions, a nonconvex problem whose global minimum is hard to certify; numerical searches give only upper estimates. By contrast, $s_{\caM}$ is a single eigenvalue, and it is tight for complete MUBs, where it equals the floor $1/(d+1)$. It therefore gives a rigorous floor certificate for specified measurements with $L\ge d+1$ in moderate dimensions.

\clearpage
\onecolumngrid
\setcounter{page}{1}
\renewcommand{\thepage}{S\arabic{page}}
\setcounter{section}{0}
\setcounter{equation}{0}
\setcounter{table}{0}
\setcounter{lemma}{0}
\setcounter{proposition}{0}
\setcounter{secnumdepth}{2}
\renewcommand{\thesection}{S\arabic{section}}
\renewcommand{\theequation}{S\arabic{equation}}
\renewcommand{\thetable}{S\arabic{table}}
\renewcommand{\theHtable}{sm.\arabic{table}}
\renewcommand{\thelemma}{S\arabic{lemma}}
\renewcommand{\theproposition}{S\arabic{proposition}}
\renewcommand{\theHproposition}{sm.\arabic{proposition}}
\renewcommand{\theHsection}{sm.\arabic{section}}
\renewcommand{\theHsubsection}{sm.\arabic{section}.\arabic{subsection}}
\renewcommand{\theHequation}{sm.\arabic{equation}}
\begin{center}
\textbf{Near-Optimal Multiparameter Pure-State Metrology with a Constant Number of Bases: Supplemental Material}
\end{center}
\begingroup
\makeatletter
\def\l@section{\@dottedtocline{1}{0em}{2.5em}}
\def\l@subsection{\@dottedtocline{2}{2.5em}{2em}}
\def\@pnumwidth{3em}
\def\@tocrmarg{4em}
\makeatother
\tableofcontents
\endgroup
\bigskip
\setlength{\tabcolsep}{9pt}
\renewcommand{\arraystretch}{1.18}
This supplement gives the proofs of the main-text and End Matter results. Its contents identify each result and its proof. Statements in the main text and End Matter retain their numbering; equations and auxiliary results introduced here carry an S prefix. Throughout, $\caH$ is the $d$-dimensional complex Hilbert space of the main text, $d\ge2$. Pure-state and state--direction kets are normalized unless explicitly stated otherwise; arbitrary tangent and Gaussian vectors need not be. Vector norms are Euclidean. Operator traces on $\caH$ and on the real physical tangent space are denoted by $\tr$ and $\tr_{\bbR }$, respectively. Zero-probability score terms are omitted.

\section{Directional Fisher information and approximate symmetry} \label{supp:normalization}
This section derives \eref{eq:basis-resp}, proves the equivalent formula for the floor in \eref{eq:fisher-floor}, and establishes the equivalence between the floor bound $\eta\ge1/2-\varepsilon$ and \eref{eq:approx-symmetry}.
\subsection{Fisher eigenvalues and directional responses} \label{supp:directional-proof}
Fix a finite POVM $\caM=\{E_i\}_{i=1}^M$ with nonzero effects on $\caH$. Define the compact set of unit state--direction pairs by
\begin{equation} \label{eq:admissible-pairs}
\caS_d:=\{(\ket\psi,\ket\nu)\in\caH\times\caH:\langle\psi|\nu\rangle=0,\ \|\ket\psi\|=\|\ket\nu\|=1\}.
\end{equation}
For $(\ket\psi,\ket\nu)\in\caS_d$, put
\begin{equation} \label{eq:general-dir-resp}
O_{\psi,\nu}:=\ket\psi\bra\nu+\ket\nu\bra\psi,\quad p_i(\psi):=\bra\psi E_i\ket\psi,\quad
g_{\caM}(\psi,\nu):=\frac{1}{4}\sum_{i:p_i(\psi)>0}\frac{[\tr(E_iO_{\psi,\nu})]^2}{p_i(\psi)}.
\end{equation}
For a basis $\caB=\{\ket{\varphi_k}\}_{k=0}^{d-1}$ of $\caH$, its effects are $\proj{\varphi_k}$, and its state and tangent amplitudes are
\begin{equation} \label{eq:amplitudes}
a_k:=\langle\varphi_k|\psi\rangle,\quad v_k:=\langle\varphi_k|\nu\rangle.
\end{equation}
The normalized directional response of this basis is
\begin{equation} \label{eq:basis-resp}
g_{\caB}(\psi,\nu):=\sum_{k:a_k\ne0}\frac{\left[\Re(\overline{a}_kv_k)\right]^2}{|a_k|^2}.
\end{equation}
More generally, fix $L$ bases $\caB_\ell=\{\ket{\varphi_{\ell k}}\}_{k=0}^{d-1}$, $\ell=0,\ldots,L-1$. The POVM $\caM$ that selects each basis with probability $1/L$ has effects
\begin{equation} \label{eq:mixture-effects}
E_{\ell k}:=\frac{\proj{\varphi_{\ell k}}}{L},\quad \ell=0,\ldots,L-1,\quad k=0,\ldots,d-1.
\end{equation}
Its response and floor obey
\begin{equation} \label{eq:dir-fisher-resp}
g_{\caM}(\psi,\nu)=\frac{1}{L}\sum_{\ell=0}^{L-1}g_{\caB_\ell}(\psi,\nu),\quad
\eta(\caM)=\inf_{(\ket\psi,\ket\nu)\in\caS_d}g_{\caM}(\psi,\nu).
\end{equation}
\begin{proof}[Derivation of \eqssref{eq:fisher-floor}{eq:basis-resp}{eq:dir-fisher-resp}]
The physical tangent space at $\ket\psi$ is represented by $\ket\psi^\perp$, viewed as a real inner-product space with inner product $\Re\langle\chi|\xi\rangle$ for $\ket\chi,\ket\xi\in\ket\psi^\perp$. A tangent ket $\ket\nu$ is horizontal when $\langle\psi|\nu\rangle=0$. For $(\ket\psi,\ket\nu)\in\caS_d$, consider
\begin{equation}
\ket{\psi(t)}:=\frac{\ket\psi+t\ket\nu}{\sqrt{1+t^2}\lsp },\quad
\left.\frac{\rmd}{\rmd t}\bigl[\proj{\psi(t)}\bigr]\right|_{t=0}=O_{\psi,\nu}.
\end{equation}
\Eref{eq:fisher-matrices} gives QFI $4$ along this curve at $t=0$ and probability derivatives $\tr(E_iO_{\psi,\nu})$. Dividing its classical Fisher information by four yields \eref{eq:general-dir-resp}. For the basis $\caB$, the derivative of $|a_k|^2$ is $2\Re(\overline{a}_kv_k)$, giving \eref{eq:basis-resp}. For the equal mixture specified in \eref{eq:mixture-effects}, each contribution is scaled by $1/L$, proving the response identity in \eref{eq:dir-fisher-resp}. The unit horizontal kets represent all tangent directions. At fixed $\psi$, the minimum of $g_{\caM}(\psi,\nu)$ over these kets is $\lambda_{\min}(G_\psi)$, giving the floor in \eref{eq:fisher-floor}.

The QFI matrix $J_\psi$ is positive definite in local coordinates. For a nonzero $\bm x\in\bbR^{2d-2}$, put $\bm y=J_\psi^{1/2}\bm x$. The directional information fraction is
\begin{equation} \label{eq:fisher-eigenvalues}
\frac{\bm x^{\sfT}I_{\caM,\psi}\bm x}{\bm x^{\sfT}J_\psi\bm x}
=\frac{\bm y^{\sfT}G_\psi\bm y}{\bm y^{\sfT}\bm y}.
\end{equation}
The stationary values are the eigenvalues of $G_\psi$, equivalently the generalized eigenvalues satisfying $I_{\caM,\psi}\bm x=\lambda J_\psi\bm x$. This proves the information-fraction interpretation in the main text. Because the two Fisher matrices undergo the same congruence under a parameter change, these eigenvalues are independent of the choice of local coordinates.

For the Cram\'er--Rao statement, let $\bm s$ be the score of $N$ independent outcomes and let $\widehat{\bm\theta}$ be a locally unbiased estimator with finite covariance. On the support of the outcome distribution, differentiation gives $\E\bm s=\bm0$, $\E\left[\bm s\bm s^{\sfT}\right]=NI_{\caM,\psi}$, and $\E\bigl[\bigl(\widehat{\bm\theta}-\bm\theta\bigr)\bm s^{\sfT}\bigr]=\bbone$. Zero-probability outcomes have zero first derivatives, so they do not alter these identities. The matrix Cauchy--Schwarz inequality gives $\Cov(\widehat{\bm\theta})\succeq(NI_{\caM,\psi})^{-1}$ whenever $I_{\caM,\psi}$ is invertible. If $I_{\caM,\psi}\succeq cJ_\psi$ with $c>0$, inversion gives $I_{\caM,\psi}^{-1}\preceq c^{-1}J_\psi^{-1}$. This bounds the inverse-information benchmark, not the covariance of an arbitrary estimator.

At a zero-probability outcome, positivity implies $E_i\ket\psi=0$ whenever $p_i(\psi)=0$, and the omitted derivative is therefore zero. For a smooth submodel, both Fisher matrices transform by congruence with the same parameter Jacobian. Restricting the tangent space preserves every lower matrix bound and may increase the minimum information fraction.
\end{proof}

At a regular state and fixed $d$, suppose $\sqrt{N}(\widehat{\bm\theta}-\bm\theta)$ converges in distribution and second moments to a centered normal vector with covariance $I_{\caM,\psi}^{-1}$. The local expansion of infidelity is $(\widehat{\bm\theta}-\bm\theta)^{\sf T}J_\psi(\widehat{\bm\theta}-\bm\theta)/4$ to second order, by \eref{eq:fisher-matrices}. Convergence in distribution and second moments justifies averaging this expansion, so the leading mean infidelity coefficient is $\tr\left(J_\psi I_{\caM,\psi}^{-1}\right)/4=\tr\left(G_\psi^{-1}\right)/4$. Under \eref{eq:approx-symmetry}, each eigenvalue of $G_\psi^{-1}$ is at most $2/(1-2\varepsilon)$. Summing over $2d-2$ tangent directions gives the bound in the Summary.

For a finite POVM with nonzero effects, Ref.~\cite[Proposition~2]{YangXiaoZhu2026} proves that the all-state floor is attained and equals the regular-state floor.
\begin{equation} \label{eq:regular-floor}
\eta_{\reg}(\caM):=\inf_{\substack{\ket\psi\in\caH\\p_i(\psi)>0\ \forall i}}\lambda_{\min}(G_\psi)=\eta(\caM).
\end{equation}

\subsection{Equivalence of approximate symmetry and a Fisher floor} \label{supp:symmetry-proof}
\begin{proof}[Equivalence of the floor bound and \eref{eq:approx-symmetry}]
For an orthonormal state--direction pair, put $p_i=\bra\psi E_i\ket\psi$. Positivity and Cauchy--Schwarz give
\begin{equation} \label{eq:phase-paired-povm}
g_{\caM}(\psi,\nu)+g_{\caM}(\psi,\rmi\nu)=\sum_{i:p_i>0}\frac{|\bra\nu E_i\ket\psi|^2}{p_i}\le\sum_{i:p_i>0}\bra\nu E_i\ket\nu\le1.
\end{equation}
Indeed, $|\bra\nu E_i\ket\psi|^2\le p_i\bra\nu E_i\ket\nu$. For a basis, this specializes to
\begin{equation} \label{eq:phase-paired-resp}
g_{\caB}(\psi,\nu)+g_{\caB}(\psi,\rmi\nu)=\sum_{k:a_k\ne0}|v_k|^2\le1.
\end{equation}
Each response is at least $\eta(\caM)$, so \eref{eq:phase-paired-povm} bounds every response above by $1-\eta(\caM)$. Hence $\eta(\caM)\ge1/2-\varepsilon$ places the spectrum of $G_\psi$ in $[1/2-\varepsilon,1/2+\varepsilon]$ at every state, including singular states. Congruence with $J_\psi^{1/2}$ gives \eref{eq:approx-symmetry}; its lower inequality proves the converse. Equivalently, $\|G_\psi-\bbone/2\|_{\op}\le\varepsilon$. The general rank-one geometry, including its trace defect at singular inputs, is discussed in Ref.~\cite[Proposition~1]{YangXiaoZhu2026}.
\end{proof}

\section{Mean Fisher anisotropy and mutually unbiased bases} \label{supp:fisher-moments}
We prove \psref{prop:mean-anisotropy}{prop:mub-floor}, derive the weighted Haar comparison, and calculate the odd-prime MUB spectrum in \pref{prop:mub-spectrum}. These results explain the distinction between average balance and the Fisher floor. Their application to basis-count limits is developed separately in \sref{supp:basis-ceilings}. All floors use the ordinary-score convention at every pure state.

\subsection{Fisher projectors and information shared by two bases} \label{supp:spec-pairing}
We represent the normalized Fisher operator on the real tangent space $\ket\psi^\perp$, with the inner product specified in SM \sref{supp:directional-proof}. The QFI identifies this representation with the coordinates in \eref{eq:fisher-floor}, without changing its spectrum. The operator $G_\psi$ and the projectors below act real-linearly; the bra--ket expression in \eref{eq:fisher-operator} is evaluated with this real inner product. For a basis $\caB=(\ket{\varphi_j})_{j=0}^{d-1}$, set $a_j:=\langle\varphi_j|\psi\rangle$ and $\ket{\chi_j}:=(a_j/|a_j|)\ket{\varphi_j}$ when $a_j\ne0$. Let $Q_{\psi,\caB}$ project orthogonally onto their real span, and let $Q_{\bbR\psi}$ project orthogonally onto the real line through $\ket\psi$. The normalized Fisher operator of this basis is
\begin{equation} \label{eq:tangent-fisher-proj}
\sfP_{\psi,\caB}:=\left.(Q_{\psi,\caB}-Q_{\bbR\psi})\right|_{\ket\psi^\perp}.
\end{equation}
Indeed, the kets $\ket{\chi_j}$ are real orthonormal, $\ket\psi=\sum_{a_j\ne0}|a_j|\ket{\chi_j}$, and $\rmi\ket\psi$ is real orthogonal to their span. For $\ket\nu\perp\ket\psi$, the squared real components of $\ket\nu$ along the $\ket{\chi_j}$ sum to $g_{\caB}(\psi,\nu)$ by \eref{eq:basis-resp}. The projector $Q_{\bbR\psi}$ annihilates $\ket\nu$, giving the quadratic form of \eref{eq:tangent-fisher-proj}. If $m_\psi$ outcomes have nonzero probability, this projector has rank $m_\psi-1$. At a regular state, this is the real-projector representation of Ref.~\cite[Lemma~3]{LuSha2026}; the formula above also specifies our zero-probability convention.

For the mixture $\caM(q)$ of the bases $\caB_0,\ldots,\caB_{L-1}$ with $q_\ell\ge0$ and $\sum_{\ell=0}^{L-1}q_\ell=1$,
\begin{equation} \label{eq:fisher-operator}
G_\psi=\sum_{\ell=0}^{L-1}q_\ell\sfP_{\psi,\caB_\ell},\quad g_{\caM}(\psi,\nu)=\Re \bra\nu G_\psi\ket\nu.
\end{equation}
Call a state regular for this family when all active basis overlaps are nonzero, as defined in the main text. The excluded states lie in a finite union of hyperplanes and have Haar measure zero. At a regular state of a basis, the real spans of $(\ket{\chi_j})$ and $(\rmi\ket{\chi_j})$ are orthogonal complements. Removing the real lines through $\ket\psi$ and $\rmi\ket\psi$, respectively, gives complementary subspaces of $\ket\psi^\perp$ interchanged by multiplication by $\rmi$. Thus the first identity below holds at a regular state of $\caB$. Summing over the active bases gives the second identity at a regular state of the mixture.
\begin{equation} \label{eq:fisher-spec-pairing}
\sfP_{\psi,\caB}(\rmi\ket\nu)=\rmi(\bbone-\sfP_{\psi,\caB})\ket\nu,\quad G_\psi(\rmi\ket\nu)=\rmi(\bbone-G_\psi)\ket\nu.
\end{equation}
Hence $\tr_{\bbR }G_\psi=d-1$, and multiplication by $\rmi$ maps each $\lambda$ eigenspace to the $1-\lambda$ eigenspace. At $\lambda=1/2$ this map squares to $-\bbone$, so the multiplicity is even. We use these conclusions at regular states. The rank formula above applies at singular states as well; in particular, $\sfP_{\psi,\caB}=0$ if $\ket\psi$ is a vector of $\caB$.

For a finite rank-one POVM with effects $E_j=w_j\proj{\chi_j}$, Ref.~\cite[Proposition~1]{YangXiaoZhu2026} gives
\begin{equation} \label{eq:rank-one-trace}
\tr_{\bbR}G_\psi=d-1-\sum_{j:p_j(\psi)=0}w_j.
\end{equation}
For Haar-almost every state, all outcome probabilities are positive and the trace is $d-1$. For a continuous rank-one POVM, the same trace calculation replaces the omitted sum by the measure of outcomes orthogonal to $\ket\psi$. Each outcome ket is orthogonal to a Haar-null set, so averaging over states and outcomes gives zero mass for the omitted outcomes at Haar-almost every state. Thus the same variance interpretation of $\wp$ applies to both classes of rank-one POVMs.

As in the main text, the mean Fisher anisotropy is
\begin{equation} \label{eq:average-deviation-def} \wp(\caM)=\frac{1}{2(d-1)}\int\tr_{\bbR}\left[\left(G_\psi-\frac{\bbone}{2}\right)^2\right]\,d\psi =\frac{1}{2(d-1)}\int\tr_{\bbR }G_\psi^2\,d\psi-\frac{1}{4}. \end{equation}
For mixtures of bases, the second equality uses the trace $d-1$ almost everywhere. Thus an average second-moment bound gives a bound on $\wp$.

Define
\begin{equation} \label{eq:moment-kernel}
f(x):=x+(1-x)\ln(1-x),\quad 0\le x\le1,
\end{equation}
with $0\ln0=0$. The following identity yields the weighted anisotropy bound in \eref{eq:weighted-deviation}.
\begin{lemma}[Pairwise Fisher moment] \label{lem:pairwise-fisher}
Let $d\ge2$, and let $\caB=(\ket{\varphi_j})_{j=0}^{d-1}$ and $\caC=(\ket{\phi_k})_{k=0}^{d-1}$ be orthonormal bases of $\caH$. Put $x_{jk}:=|\langle\varphi_j|\phi_k\rangle|^2$. Then
\begin{equation} \label{eq:pairwise-fisher-moment}
\int\tr_{\bbR }(\sfP_{\psi,\caB}\sfP_{\psi,\caC})\,d\psi=\frac{d-1}{2}+\frac{1}{2}\left[\sum_{j,k}f(x_{jk})-1\right].
\end{equation}
\end{lemma}
\begin{proof}[Proof of \lref{lem:pairwise-fisher}]
For almost every $\psi$, all amplitudes $a_j:=\langle\varphi_j|\psi\rangle$ and $\widetilde{a}_k=\langle \phi_k|\psi\rangle$ are nonzero. The ranges of both ambient projectors contain $\ket\psi$, so \eref{eq:tangent-fisher-proj} gives
\begin{equation} \label{eq:moment-proj-trace}
\tr_{\bbR }(\sfP_{\psi,\caB}\sfP_{\psi,\caC})=\sum_{j,k}\left[\Re \left(\langle\varphi_j|\phi_k\rangle\frac{\overline{a}_j\widetilde{a}_k}{|a_j\widetilde{a}_k|}\right)\right]^2-1.
\end{equation}
Fix one summand and put $x=x_{jk}$. Rephase its basis vectors so that their overlap is $\sqrt{x}$. Represent the Haar state by a normalized standard complex Gaussian vector~\cite{Mezzadri2007}; its norm cancels from the phases. The relevant amplitudes therefore have the phases of $z$ and $w:=z\sqrt{x}\lsp +y\sqrt{1-x}$, where $z,y$ are independent standard complex Gaussians.

For $0<x<1$, condition on $|z|,|y|$ and define $\kappa\ge0$ by $\kappa^2=(1-x)|y|^2/(x|z|^2)$. The relative phase $\vartheta:=\arg y-\arg z$ is uniform on $[0,2\pi)$, and $\arg w-\arg z=\arg\left(1+\kappa\rme^{\rmi\vartheta}\right)$ modulo $2\pi$. Its doubled-phase average is
\begin{equation} \label{eq:phase-angular-average}
\E _\vartheta\rme^{2\rmi(\arg w-\arg z)}=\frac{1}{2\pi}\int_0^{2\pi}\frac{1+\kappa\rme^{\rmi\vartheta}}{1+\kappa\rme^{-\rmi\vartheta}}\,\rmd \vartheta=\max\left\{1-\kappa^2,0\right\}.
\end{equation}
For $\kappa<1$, expanding the reciprocal denominator as a uniformly convergent geometric series gives constant term $1-\kappa^2$ for the quotient. For $\kappa>1$, rewrite the quotient as $\rme^{\rmi\vartheta}\left(\kappa^{-1}+\rme^{\rmi\vartheta}\right)/(1+\kappa^{-1}\rme^{\rmi\vartheta})$ and expand in $\kappa^{-1}$; all powers of $\rme^{\rmi\vartheta}$ are positive, so their averages vanish. At $\kappa=1$, the quotient equals $\rme^{\rmi\vartheta}$ except at one angle, and its average is again zero.

The squared moduli are independent unit-rate exponentials. Hence $u:=|z|^2/(|z|^2+|y|^2)$ is uniform on $(0,1)$ by the change of variables from the two exponentials to their sum and ratio. Since $\kappa^2=(1-x)(1-u)/(xu)$, the angular average is positive precisely when $u>1-x$. Taking real parts in \eref{eq:phase-angular-average} and using the double-angle identity for $\cos^2$ gives
\begin{equation} \label{eq:phase-square-moment} \E \left\{\left[\Re \left(\sqrt{x}\lsp \frac{\overline{z}w}{|zw|}\right)\right]^2\right\} =\frac{x}{2}+\frac{1}{2}\int_{1-x}^1\left(1-\frac{1-x}{u}\right)\,\rmd u =\frac{x+f(x)}{2}. \end{equation}
At $x=0$ the summand is zero; at $x=1$, $w=z$ and the summand is one. Both values agree with \eref{eq:phase-square-moment}. Sum this identity in \eref{eq:moment-proj-trace} and use $\sum_{j,k}x_{jk}=d$ to obtain \eref{eq:pairwise-fisher-moment}. Zero amplitudes occur only on a set of Haar measure zero, and the projectors are bounded, so they do not affect the integral.
\end{proof}

\subsection{Proof of \texorpdfstring{\pref{prop:mean-anisotropy}}{Proposition 1} and its weighted extension} \label{supp:average-mub}
\begin{proof}[Proof of \pref{prop:mean-anisotropy} and its weighted extension]
Let $\caM(q)$ be the POVM that selects basis $\caB_\ell$ with probability $q_\ell>0$, where $\sum_{\ell=0}^{L-1}q_\ell=1$, and let $\varsigma:=\sum_\ell q_\ell^2$. Expanding \eref{eq:fisher-operator} and applying \lref{lem:pairwise-fisher} gives the exact identity
\begin{equation} \label{eq:exact-fisher-second}
 4\wp(\caM(q))=\varsigma+\frac{2}{d-1}\sum_{\ell<\ell'}q_\ell q_{\ell'}\left[\sum_{j,k}f\left(|\langle\varphi_{\ell j}|\varphi_{\ell' k}\rangle|^2\right)-1\right].
\end{equation}
The continuous function $f$ in \eref{eq:moment-kernel} is strictly convex: $f''(x)=1/(1-x)>0$ on $[0,1)$. Since the $d^2$ squared overlaps of two bases sum to $d$, Jensen's inequality yields
\begin{equation} \label{eq:overlap-jensen}
 \sum_{j,k}f\left(|\langle\varphi_{\ell j}|\varphi_{\ell' k}\rangle|^2\right)\ge d^2f\left(\frac{1}{d}\right),
\end{equation}
with equality exactly when every squared overlap is $1/d$. Using $2\sum_{\ell<\ell'}q_\ell q_{\ell'}=1-\varsigma$ in \eref{eq:exact-fisher-second}, we obtain
\begin{equation} \label{eq:weighted-deviation}
 4\wp(\caM(q))\ge \varsigma+\frac{1-\varsigma}{d-1}\left[d^2f\left(\frac{1}{d}\right)-1\right]
 =1-(1-\varsigma)d\ln\frac{d}{d-1}.
\end{equation}
This proves the weighted bound used in \pref{prop:basis-ceiling}. By $\varsigma\ge1/L$, the bound is minimized at equal weights. For $L\ge2$ and positive setting weights, equality in \eref{eq:weighted-deviation} holds exactly for pairwise mutually unbiased bases. For $L=1$, the Fisher operator is a rank-$(d-1)$ projector almost everywhere and $\wp=1/4$. If some weights vanish, apply the statement to the retained settings. The proof of \pref{prop:basis-ceiling} combines this moment bound with spectral pairing to obtain \eqsref{eq:finite-settings}{eq:fixed-settings-ceiling}.

For equal weights, \eref{eq:weighted-deviation} is precisely \eref{eq:average-deviation}, proving \pref{prop:mean-anisotropy} with its equality condition.
For a complete MUB family with equal weights, equality in \eref{eq:average-deviation} gives
\begin{equation} \label{eq:complete-mub-average}
 \wp(\caM_{\MUB})=\frac{1}{4}\left[1-\frac{d^2}{d+1}\ln\frac{d}{d-1}\right]
 =\frac{1}{8d}+\caO\left(d^{-2}\right).
\end{equation}
The asymptotic expansion is along dimensions admitting complete families and follows from $d\ln[d/(d-1)]=1+1/(2d)+\caO\left(d^{-2}\right)$.
The weighted bound also gives a finite-dimensional necessary setting count. If $\eta(\caM)\ge1/2-\varepsilon$, \eref{eq:approx-symmetry} gives $\wp\le\varepsilon^2$. Combining this with $\varsigma\ge1/L$ in \eref{eq:weighted-deviation} yields
\begin{equation} \label{eq:finite-resource}
L\ge\frac{d\ln\frac{d}{d-1}}{4\varepsilon^2+d\ln\frac{d}{d-1}-1}.
\end{equation}
The denominator is positive because $d\ln[d/(d-1)]>1$. This proves the finite-dimensional counterpart of \eref{eq:fixed-settings-ceiling}.

\end{proof}

\subsection{Mean Fisher anisotropy of weighted Haar bases} \label{supp:haar-anisotropy}
The following proposition proves the Haar average stated in the main text and extends it to arbitrary allocation probabilities.

\begin{proposition}[Fisher averages for weighted Haar bases] \label{prop:haar-anisotropy}
	Let $\caB_0,\ldots,\caB_{L-1}$ be independent Haar orthonormal bases of $\caH$, with $L\ge1$. Let $\caM(q)$ be the POVM that selects basis $\caB_\ell$ with fixed probability $q_\ell\ge0$, where $\sum_{\ell=0}^{L-1}q_\ell=1$, and let $\varsigma:=\sum_{\ell=0}^{L-1}q_\ell^2$. For each fixed pure state $\ket\psi\in\caH$, averaging over the bases gives
	\begin{equation} \label{eq:haar-weighted}
		\E G_\psi=\frac{\bbone}{2},\quad
		\frac{1}{2(d-1)}\E\tr_{\bbR}\left[\left(G_\psi-\frac{\bbone}{2}\right)^2\right]
		=\E\wp(\caM(q))=\frac{\varsigma}{4}.
	\end{equation}
	For fixed $L$, the expected mean Fisher anisotropy $\E\wp(\caM(q))$ is uniquely minimized by $q_\ell=1/L$ for all $\ell$.	If $L$ mutually unbiased bases exist, let $\caM_{\MUB}(q)$ be the POVM that mixes them with the same probabilities. Then
	\begin{equation} \label{eq:haar-mub-gap}
		\E\wp(\caM(q))-\wp(\caM_{\MUB}(q))
		=\frac{1-\varsigma}{4}\left(d\ln\frac{d}{d-1}-1\right).
	\end{equation}
\end{proposition}

\begin{proof}[Proof of \pref{prop:haar-anisotropy}]
At a fixed $\psi$, the single-basis Fisher operator $\sfP_\ell$ is almost surely a rank-$(d-1)$ real orthogonal projector by \eref{eq:tangent-fisher-proj}. The Beta distribution in \lref{lem:beta-laplace} gives mean directional response $1/2$, so polarization yields $\E \sfP_\ell=\bbone/2$. Hence $\E G_\psi=\bbone/2$. Put $A_\ell=\sfP_\ell-\bbone/2$. Then $A_\ell^2=\bbone/4$ and $\E A_\ell=0$. Independence makes every expected cross trace vanish, giving
\begin{equation}
\E\tr_{\bbR}\left[\left(\sum_\ell q_\ell A_\ell\right)^2\right]
=\frac{2(d-1)}{4}\sum_\ell q_\ell^2.
\end{equation}
Division by $2(d-1)$ proves the fixed-input identity. Integrating over normalized Haar-state measure proves the identity for $\wp$; boundedness justifies interchanging the expectations. Cauchy--Schwarz gives $\varsigma\ge1/L$, with equality exactly for equal probabilities. Finally, subtract the equality value in \eref{eq:weighted-deviation} from $\varsigma/4$ to obtain \eref{eq:haar-mub-gap}. Zero probabilities are handled by omitting those bases.
\end{proof}

\subsection{Proof of \texorpdfstring{\pref{prop:mub-floor}}{Proposition 2}: complete MUBs and subsets} \label{supp:mub-benchmark}
\begin{proof}[Proof of \pref{prop:mub-floor}] Write $\caB_\ell=\{\ket{\varphi_{\ell k}}\}_{k=0}^{d-1}$ for each basis in the proposition. For orthonormal $\ket\psi,\ket\nu\in\caH$, put $O=\ket\psi\bra\nu+\ket\nu\bra\psi$, $P_{\ell k}=\ket{\varphi_{\ell k}}\bra{\varphi_{\ell k}}$, $p_{\ell k}=\bra\psi P_{\ell k}\ket\psi$, and $x_{\ell k}=\tr(P_{\ell k}O)$. In each basis, $\sum_kx_{\ell k}=0$ and $p_{\ell k}=0$ implies $x_{\ell k}=0$.
\begin{equation} \label{eq:single-basis-quadratic}
 \sum_{k:p_{\ell k}>0}\frac{x_{\ell k}^2}{p_{\ell k}}
 \ge\left(\sum_k|x_{\ell k}|\right)^2
 \ge2\sum_kx_{\ell k}^2.
\end{equation}
Cauchy--Schwarz gives the first inequality because the probabilities in each basis sum to one. For the second, the positive and negative entries of $(x_{\ell k})_k$ each have total absolute value $\frac12\sum_k|x_{\ell k}|$, so their total sum of squares is at most $2(\frac12\sum_k|x_{\ell k}|)^2$. The traceless diagonal subspaces of the $d+1$ bases are mutually orthogonal in the Hilbert--Schmidt inner product. Their dimensions sum to $d^2-1$, so they span all traceless Hermitian operators; this is also the 2-design identity for a complete MUB family~\cite{KlappeneckerRoetteler2005,RoyScott2007}. The squared Hilbert--Schmidt norm of the diagonal projection of $O$ in basis $\ell$ is $\sum_k x_{\ell k}^2$. Orthogonality and completeness give $\sum_{\ell,k}x_{\ell k}^2=\tr\left(O^2\right)=2$. Using \eref{eq:single-basis-quadratic} in the directional response \eref{eq:general-dir-resp} gives $g_{\caM_{\MUB}(q)}(\psi,\nu)\ge\min_\ell q_\ell$. To attain this value, choose $\ell_*$ minimizing $q_\ell$ and put
\begin{equation}
\ket\psi:=\frac{\ket{\varphi_{\ell_*0}}+\ket{\varphi_{\ell_*1}}}{\sqrt{2}\lsp },\quad
\ket\nu:=\frac{\ket{\varphi_{\ell_*0}}-\ket{\varphi_{\ell_*1}}}{\sqrt{2}\lsp },\quad
O=\proj{\varphi_{\ell_*0}}-\proj{\varphi_{\ell_*1}}.
\end{equation} That basis has normalized response one; mutual unbiasedness makes all other derivatives zero. The mixture response is its weight, proving the formula also for omitted bases. For $d=2$, take the three Pauli eigenbases
\begin{equation} \label{eq:pauli-bases}
\caB_x=\left\{\frac{\ket0\pm\ket1}{\sqrt{2}\lsp }\right\},\quad
\caB_y=\left\{\frac{\ket0\pm\rmi\ket1}{\sqrt{2}\lsp }\right\},\quad
\caB_z=\{\ket0,\ket1\}.
\end{equation}
Their floor is exactly $\min\{q_x,q_y,q_z\}$. Equal probabilities give $1/3$, saturating the $1/L$ ceiling in \eref{eq:finite-ceiling} for $L=3$.

\end{proof}

\subsection{Proof of \texorpdfstring{\pref{prop:mub-spectrum}}{Proposition 5}: Fisher spectrum for odd-prime MUBs} \label{supp:mub-spectrum}
Let $d$ be an odd prime, let $\bbF_d:=\bbZ/d\bbZ$ be the field of integers modulo $d$, and put $\omega=\rme^{2\pi\rmi/d}$. Alongside the computational basis, take the $d$ quadratic-phase bases
\begin{equation} \label{eq:prime-mub-bases}
\ket{\varphi_{\ell k}}:=\frac{1}{\sqrt{d}\lsp }\sum_{j\in\bbF _d}\omega^{\ell j(j-1)+kj}|j\rangle,\quad \ell,k\in\bbF _d.
\end{equation}
Together with $\caB_d:=\{\ket j\}_{j=0}^{d-1}$, these form the complete MUB family in \pref{prop:mub-spectrum}~\cite{WoottersFields1989}. Let $\caM_{\MUB}$ be the POVM uniformly mixing these $d+1$ bases. We compute its spectrum at $\ket{\psi_+}:=(\ket0+\ket1)/\sqrt{2}$.

\begin{proof}[Proof of \pref{prop:mub-spectrum}]
Write an arbitrary horizontal tangent as $\ket{\dot\psi}=(x+\rmi y)\ket{\nu_0}+\ket\chi$, where $\ket{\nu_0}=(\ket0-\ket1)/\sqrt{2}$, $x,y\in\bbR$, and $\ket\chi=\sum_{j\ne0,1}z_j\ket j$. Extend these coefficients by $z_0=-z_1=(x+\rmi y)/\sqrt{2}$, so $\ket{\dot\psi}=\sum_{j\in\bbF_d}z_j\ket j$. For the quadratic-phase bases in \eref{eq:prime-mub-bases}, the state and tangent amplitudes are
\begin{equation} \label{eq:mub-spectrum}
a_{\ell k}:=\langle\varphi_{\ell k}|\psi_+\rangle=\frac{1+\omega^{-k}}{\sqrt{2d}\lsp },\quad
v_{\ell k}:=\langle\varphi_{\ell k}|\dot\psi\rangle=\frac{1}{\sqrt{d}\lsp }\sum_{j\in\bbF_d}z_j\omega^{-\ell j(j-1)-kj}.
\end{equation}
Since $d$ is odd, $a_{\ell k}\ne0$ and $\overline{a}_{\ell k}/a_{\ell k}=\omega^k$. Write $I$ and $J$ as quadratic forms on the physical tangent space. Then \eref{eq:basis-resp}, homogeneity in the tangent, and $(\Re w)^2=\left(|w|^2+\Re w^2\right)/2$ give
\begin{equation}
I_{\caB_\ell,\psi_+}(\dot\psi,\dot\psi)
=2\norm{\ket{\dot\psi}}^2+2\Re\sum_k\omega^kv_{\ell k}^2.
\end{equation}
Character orthogonality yields
\begin{equation} \label{eq:mub-character-sums}
\sum_{\ell,k}\omega^kv_{\ell k}^2
=d\sum_{\substack{j+j'=1\\j(j-1)+j'(j'-1)=0}}z_jz_{j'}
=2dz_0z_1=-d(x+\rmi y)^2.
\end{equation}
The constraints are in $\bbF_d$. Substituting $j'=1-j$ into the second gives $2j(j-1)=0$, so only $(j,j')=(0,1),(1,0)$ remain. Consequently, the quadratic-phase bases contribute $4dy^2+2d\norm{\ket\chi}^2$ in total. The computational basis contributes $4x^2$, because only its probabilities at 0 and 1 are nonzero and their derivatives are $x,-x$. Equal mixing therefore gives
\begin{equation} \label{eq:prime-mub-fisher}
I_{\caM_{\MUB},\psi_+}(\dot\psi,\dot\psi)=\frac{4x^2+4dy^2+2d\norm{\ket\chi}^2}{d+1},\quad
J_{\psi_+}(\dot\psi,\dot\psi)=4\left(x^2+y^2+\norm{\ket\chi}^2\right).
\end{equation}
The $x$ direction, the $2d-4$ real directions in $\ket\chi$, and the $y$ direction have eigenvalues $1/(d+1)$, $d/[2(d+1)]$, and $d/(d+1)$, respectively. This proves the stated spectrum and multiplicities, including the zero-probability computational outcomes through the omission convention in \eref{eq:general-dir-resp}.
\end{proof}

\section{Probability tools for establishing constant Fisher floors} \label{supp:common-tools}
\thsref{thm:large-floors}{thm:near-half} combine the fixed-pair tail estimates below with a Haar-cap volume bound to obtain uniform floors. \Thref{thm:five-bases} uses the rank-two argument explained in \sref{supp:five-bases}. The truncated responses use the amplitudes in \eref{eq:amplitudes}, and each equal mixture has the effects in \eref{eq:mixture-effects}.

For $\tau>0$, set
\begin{equation} \label{eq:scalar-trunc}
h_\tau(a,v):=\frac{\Re (\overline{a}v)}{\max\left\{|a|,\sqrt{\tau}\lsp |v|\right\}},\quad h_\tau(0,0):=0.
\end{equation}
The nonnegative truncated responses are defined by
\begin{equation} \label{eq:trunc-mixture-resp}
R_{\tau,\caB}^2(\psi,\nu):=\sum_k h_\tau^2(a_k,v_k),\quad
R_{\tau,\caM}^2(\psi,\nu):=\frac{1}{L}\sum_{\ell=0}^{L-1} R_{\tau,\caB_\ell}^2(\psi,\nu).
\end{equation}
Unlike $g_{\caM}$, these truncated responses are continuous at zero amplitudes, as proved in \lref{lem:common-trunc}.

\subsection{Probability preliminaries} \label{supp:qr-preliminaries}

A standard circular complex Gaussian $z\sim\caN_{\bbC }(0,1)$ has independent real and imaginary parts of variance $1/2$; hence $|z|^2\sim\Exp(1)$. For shape $\alpha>0$ and scale $\beta>0$, we use the shape--scale convention $\Gamma(\alpha,\beta)$ with density
\begin{equation} \label{eq:gamma-density}
 f_{\alpha,\beta}(x):=\frac{x^{\alpha-1}\rme^{-x/\beta}}{\Gamma(\alpha)\beta^\alpha},\quad x>0.
\end{equation}
Its mean is $\alpha\beta$ and its variance is $\alpha\beta^2$. We use $\Exp(\lambda)$ for the exponential distribution of rate $\lambda$, and $\chi_N^2$ for the distribution $\Gamma(N/2,2)$.
For $G_\alpha\sim\Gamma(\alpha,1)$, define
\begin{equation} \label{eq:gamma-cdf}
F_\alpha(y):=\Pr\{G_\alpha\le y\}=\frac{1}{\Gamma(\alpha)}\int_0^y u^{\alpha-1}\rme^{-u}\,\rmd u,\quad
\Lambda(x):=x-1-\ln x,\quad x>0.
\end{equation}

\begin{lemma}[Gamma and chi-square bounds] \label{lem:gamma-chi-bounds}
For $\alpha,y>0$,
\begin{equation} \label{eq:gamma-bounds}
F_\alpha(\alpha)\ge\frac{1}{2},\quad F_\alpha(y)\ge\frac{y^\alpha\rme^{-y}}{\Gamma(\alpha+1)}.
\end{equation}
For $Y\sim\chi_N^2$, $N\ge1$, and $0<x<1$,
\begin{equation} \label{eq:chi-square-tail}
\Pr\{Y\le Nx\}\le\exp\left[-\frac{N}{2}\Lambda(x)\right],\quad \Lambda(x)\ge\frac{(1-x)^2}{2}.
\end{equation}
\end{lemma}
\begin{lemma}[Weighted Gamma lower tails] \label{lem:weighted-gamma-mean}
Let $(G_j)_j$ be a finite independent family with $G_j\sim\Gamma(n_j,1)$ and positive integer shapes $n_j$, and let $w_j\ge0$ be not all zero. Set $N=\sum_jn_j$ and $m=\sum_jw_jn_j$. For $x\ge0$,
\begin{equation} \label{eq:weighted-gamma-mean}
 \Pr\left\{\sum_jw_jG_j\le x\right\}\ge F_N\left(N\min\left(\frac{x}{m},1\right)\right).
\end{equation}
In particular, at $x=m$ the probability is at least $F_N(N)\ge1/2$.
\end{lemma}
\begin{lemma}[Gaussian QR decomposition] \label{lem:gauss-qr}
Let $Z$ be a $d\times2$ matrix of independent standard circular complex Gaussians. Its thin QR decomposition with positive diagonal can be written
\begin{equation} \label{eq:gauss-qr}
\frac{Z}{\sqrt{d}\lsp }=Q\begin{pmatrix}r&z\\0&s\end{pmatrix},\quad
dr^2\sim\Gamma(d,1),\quad ds^2\sim\Gamma(d-1,1),\quad z\sim\caN_{\bbC }\left(0,\frac{1}{d}\right).
\end{equation}
The columns of $Q$ form a Haar two-frame, meaning an orthonormal pair with the unitary-invariant distribution. This frame is independent of the mutually independent variables $r,s,z$.
\end{lemma}
\begin{proof}[Proof of \lref{lem:gamma-chi-bounds}]
We first prove the two inequalities in \eref{eq:gamma-bounds}. The change of variable $W=\ln(G_\alpha/\alpha)$ gives a density proportional to $\exp\left[\alpha w-\alpha\rme^w\right]$. Consequently,
\begin{equation}
 \frac{f_W(-u)}{f_W(u)}=\exp\left[\alpha\left(\rme^u-\rme^{-u}-2u\right)\right]\ge1,\quad u\ge0.
\end{equation}
Integration over $u\ge0$ shows $\Pr\{W\le0\}\ge\Pr\{W\ge0\}$, hence $F_\alpha(\alpha)\ge1/2$. For the second inequality, \eref{eq:gamma-cdf} gives
\begin{equation}
 F_\alpha(y)\ge\frac{\rme^{-y}}{\Gamma(\alpha)}\int_0^y u^{\alpha-1}\,\rmd u
 =\frac{y^\alpha\rme^{-y}}{\Gamma(\alpha+1)}.
\end{equation}
To prove the tail bound in \eref{eq:chi-square-tail}, use the shape--scale convention in \eref{eq:gamma-density}. For $t>0$, exponential Markov yields
\begin{equation}
 \Pr\{Y\le Nx\}\le\rme^{tNx}\E \rme^{-tY}
 =\rme^{tNx}(1+2t)^{-N/2}.
\end{equation}
Choosing $t=(1-x)/(2x)>0$ gives the stated exponent. Finally,
\begin{equation}
 \Lambda(x)=\int_x^1\frac{1-u}{u}\,\rmd u\ge\int_x^1(1-u)\,\rmd u=\frac{(1-x)^2}{2},
\end{equation}
completing \eref{eq:chi-square-tail}.
\end{proof}

\begin{proof}[Proof of \lref{lem:weighted-gamma-mean}]
We prove \eref{eq:weighted-gamma-mean}. Write each $G_j$ as a sum of $n_j$ independent $\Exp(1)$ variables. Repeating $w_j$ exactly $n_j$ times gives coefficients $c_1,\ldots,c_N\ge0$ such that
\begin{equation} \label{eq:gamma-expanded-sum}
 \sum_jw_jG_j\overset{\rm law}=\sum_{i=1}^N c_i y_i,\quad \sum_{i=1}^N c_i=m,
\end{equation}
where the $y_i$ are independent unit-mean exponentials. For $N=1$ the claim follows directly from the exponential CDF. For $N\ge2$ and $0<x<m$, the equal-weight extremum theorem \cite[Theorem~2]{RoostaKhorasaniSzekelyAscher2015} gives
\begin{equation} \label{eq:gamma-equal-weight}
 \Pr\left\{\sum_{i=1}^N\frac{c_i}{m}y_i\le\frac{x}{m}\right\}
 \ge\Pr\left\{\frac{1}{N}\sum_{i=1}^N y_i\le\frac{x}{m}\right\}
 =F_N\left(\frac{Nx}{m}\right).
\end{equation}
The cited theorem applies below the common mean $1$, including zero coefficients. Continuity extends the bound to $x=0,m$, and monotonicity gives $F_N(N)$ as a lower bound for $x\ge m$. \Eref{eq:gamma-bounds} proves the final assertion.
\end{proof}

\begin{proof}[Proof of \lref{lem:gauss-qr}]
To establish \eref{eq:gauss-qr} and the stated independence, write the columns of $Z$ as $\bm g_1,\bm g_2\in\bbC^d$ and put $Q_1=\bm g_1/\|\bm g_1\|$ and $\bm g_\perp:=\bigl(\bbone-Q_1Q_1^\dag\bigr)\bm g_2$. Gram--Schmidt gives
\begin{equation}
 r=\frac{\|\bm g_1\|}{\sqrt{d}\lsp },\quad z=\frac{Q_1^\dag\bm g_2}{\sqrt{d}\lsp },\quad
 s=\frac{\|\bm g_\perp\|}{\sqrt{d}\lsp },\quad Q_2=\frac{\bm g_\perp}{\|\bm g_\perp\|}.
\end{equation} Conditional on $Q_1$, the parallel scalar and the perpendicular Gaussian vector are independent standard complex Gaussians on their respective subspaces. The perpendicular norm is independent of its direction $Q_2$, and the conditional distributions of $z$ and $s$ do not depend on $Q_1$. The radius of $\bm g_1$ is independent of its direction and of $\bm g_2$. These facts give all three scalar laws and their independence from the Haar frame \cite{Mezzadri2007}.
\end{proof}

\subsection{Basic properties of Gaussian and Haar responses} \label{supp:gauss-haar}

\begin{lemma}[Scaling and Gaussian response] \label{lem:resp-scaling}
For $\tau,r,s>0$ and $a,v\in\bbC $,
\begin{equation} \label{eq:adaptive-trunc}
h_\tau(ra,rv)=rh_\tau(a,v),\quad h_{\frac{\tau r^2}{s^2}}(ra,sv)=s h_\tau(a,v).
\end{equation}
For independent $z_1,z_2\sim\caN_{\bbC }(0,1)$,
\begin{equation} \label{eq:scalar-gauss-laws}
2(1+\tau)h_\tau^2(z_1,z_2)\sim\chi_1^2,\quad \E  h_\tau^2(z_1,z_2)=\frac{1}{2(1+\tau)}.
\end{equation}
\end{lemma}
\begin{lemma}[Haar row response] \label{lem:gauss-haar-means}
Let $\tau>0$.
For a fixed $(\ket\psi,\ket\nu)\in\caS_d$ and a Haar orthonormal basis $\caB=\{\ket{\varphi_k}\}_{k=0}^{d-1}$ of $\caH$, each amplitude row in \eref{eq:amplitudes} satisfies
\begin{equation} \label{eq:haar-row-laws}
(1+\tau)h_\tau^2(a_k,v_k)\sim\Beta \left(\frac{1}{2},d-\frac{1}{2}\right),\quad
\E  h_\tau^2(a_k,v_k)=\frac{1}{2d(1+\tau)},\quad \E  R_{\tau,\caB}^2=\frac{1}{2(1+\tau)}.
\end{equation}
\end{lemma}
\begin{lemma}[Symmetric perturbation] \label{lem:sym-pert}
For $\tau>0$ and $a,v,z\in\bbC $,
\begin{equation} \label{eq:sym-pert}
\frac{h_\tau^2(a,v+z)+h_\tau^2(a,v-z)}{2}\le h_\tau^2(a,v)+|z|^2.
\end{equation}
\end{lemma}

\begin{proof}[Proof of \lref{lem:resp-scaling}]
We first prove \eref{eq:adaptive-trunc}. The scaling identities follow directly from \eref{eq:scalar-trunc}. If $\theta=\arg v-\arg a$, then $h_\tau^2(a,v)=\min\left\{|v|^2,|a|^2/\tau\right\}\cos^2\theta$, with zero value at zero inputs. For the Gaussian pair, independence of the squared moduli gives, for $x\ge0$,
\begin{equation}
 \Pr\left\{\min\left(|z_2|^2,\frac{|z_1|^2}{\tau}\right)>x\right\}
 =\rme^{-x}\rme^{-\tau x}=\rme^{-(1+\tau)x}.
\end{equation}
Thus the minimum is exponential with rate $1+\tau$, independently of the uniform relative angle $\theta$. Consequently, the product of this minimum and $\cos^2\theta$ is distributed as the squared real part of a circular Gaussian of variance $1/(1+\tau)$, proving \eref{eq:scalar-gauss-laws}.

The mean of the $\chi_1^2$ distribution is one, giving the mean in \eref{eq:scalar-gauss-laws}.
\end{proof}

\begin{proof}[Proof of \lref{lem:gauss-haar-means}]
We prove \eref{eq:haar-row-laws}. In orthonormal coordinates, let $\bm z=(z_1,\ldots,z_d)\in\bbC^d$ have independent standard complex Gaussian entries. A row of a Haar unitary is distributed as $\bm z/\|\bm z\|$. Set
\begin{equation} \label{eq:haar-row-variable}
 y:=(1+\tau)h_\tau^2\left(\frac{z_1}{\|\bm z\|},\frac{z_2}{\|\bm z\|}\right).
\end{equation}
The Gaussian direction is independent of its norm. By the scaling identity in \eref{eq:adaptive-trunc} and the distribution in \eref{eq:scalar-gauss-laws},
\begin{equation} \label{eq:haar-row-product}
 \|\bm z\|^2\sim\Gamma(d,1),\quad
 \|\bm z\|^2y=(1+\tau)h_\tau^2(z_1,z_2)\sim\Gamma\left(\frac{1}{2},1\right),
\end{equation}
with $y$ independent of $\|\bm z\|$. Taking moments of the product therefore gives
\begin{equation} \label{eq:haar-row-moments}
 \E  y^m=\frac{\E \left(\|\bm z\|^2y\right)^m}{\E \|\bm z\|^{2m}}
 =\frac{(1/2)_m}{(d)_m},\quad (b)_m:=\frac{\Gamma(b+m)}{\Gamma(b)},\quad m=0,1,\ldots.
\end{equation}
For normalized coordinates $a=z_1/\|\bm z\|$ and $v=z_2/\|\bm z\|$, the scalar formula gives
\begin{equation}
0\le y\le(1+\tau)\min\left\{|v|^2,\frac{|a|^2}{\tau}\right\}\le |a|^2+|v|^2\le1.
\end{equation}
Thus the moments in \eref{eq:haar-row-moments} determine the distribution on $[0,1]$ and identify it as $\Beta (1/2,d-1/2)$. Its mean is $1/(2d)$; division by $1+\tau$ gives the row mean, and summing over the $d$ rows gives the response mean in \eref{eq:haar-row-laws}.
\end{proof}

\begin{proof}[Proof of \lref{lem:sym-pert}]
To prove \eref{eq:sym-pert}, it suffices to show that $f_a(v):=h_\tau^2(a,v)-|v|^2$ is concave on $\bbC\simeq\bbR^2$. The case $a=0$ is immediate. Otherwise, a common phase rotation of $a,v,z$ reduces to $a>0$. Put $r_*=a/\sqrt{\tau}\lsp $ and write $v=x+\rmi y$. We use the representation
\begin{equation}
f_a(v)=\min\left\{-y^2,\ \inf_{|t|\le1}\left[r_*^2-2r_*ty-\left(1-t^2\right)|v|^2\right]\right\}.
\end{equation}
Indeed, for $v\ne0$, the unrestricted quadratic minimum over $t\in\bbR$ is attained at $t=r_*y/|v|^2$ and equals $r_*^2x^2/|v|^2-|v|^2$. When $|v|\ge r_*$, this minimizer lies in $[-1,1]$ and the minimum is at most $-y^2$, so the second branch agrees with $f_a(v)$. When $0<|v|\le r_*$, the unrestricted minimum is at least $-y^2$, so the first branch applies; $v=0$ is immediate. For $|t|\le1$, the coefficient of $|v|^2$ is nonpositive, so every quadratic in the displayed infimum is concave in $(x,y)$. The function $-y^2$ is also concave, and the pointwise infimum of concave functions is concave. Thus $f_a$ is concave. Its midpoint inequality, together with $\left(|v+z|^2+|v-z|^2\right)/2=|v|^2+|z|^2$, proves \eref{eq:sym-pert}.
\end{proof}

\subsection{Tail bounds for the Fisher response and Fisher floor} \label{supp:common-statements}
A standard route from a fixed state--direction estimate to a uniform bound is to cover all admissible pairs by a finite net and apply a union bound. The untruncated Fisher response need not have a uniform continuity bound near zero probabilities. We instead work with a continuous truncated response and average the volume of a Haar cap around a putative bad pair, removing the irrelevant common phase of the pair first. This avoids constructing a discrete net and retains an explicit finite-dimensional volume factor. The lemmas below quantify the truncation loss, cap volume, and fixed-pair tails separately.

\begin{lemma}[Truncated response] \label{lem:common-trunc}
Let $\caB_\ell=\{\ket{\varphi_{\ell k}}\}_{k=0}^{d-1}$, $\ell=0,\ldots,L-1$, be orthonormal bases of $\caH$, and let $\caM$ be their equal mixture as in \eref{eq:mixture-effects}. For $\tau>0$, the function $R_{\tau,\caM}$ is continuous on $\caS_d$ and satisfies $R_{\tau,\caM}^2\le g_{\caM}$, with the same zero set. For admissible pairs,
\begin{equation} \label{eq:resp-lipschitz}
|R_{\tau,\caM}(\psi,\nu)-R_{\tau,\caM}(\psi',\nu')|
\le\tau^{-1/2}\|\ket\psi-\ket{\psi'}\|+\|\ket\nu-\ket{\nu'}\|.
\end{equation}
\end{lemma}
For $\tau>0$ and $0<\delta\le1/10$, choose the cap radii $x,y$ below and define $\zeta$ and $K_{\rm cov}$ by
\begin{equation} \label{eq:covering-cost}
x=\frac{\delta\sqrt{\tau}\lsp }{2\bigl(1+\sqrt{\tau}\lsp \bigr)},\quad y=\frac{\delta}{2},\quad
\zeta(x):=x\sqrt{1-\frac{x^2}{4}},\quad
K_{\rm cov}(\tau,\delta):=-2\ln[\zeta(x)\zeta(y)]+\delta^4.
\end{equation}
Here $x,y$ are cap radii; the function $\zeta$ is defined on $[0,1)$ and converts a chordal radius to the radius of its orthogonal projection. For real $u\ge2$, define the following smooth extension of the cap cost (at integer $u=d$, $u$ is the complex Hilbert-space dimension):
\begin{equation} \label{eq:finite-cap-cost}
 K_u(\tau,\delta):=-(2u-2)\ln\zeta(x)-(2u-3)\ln\zeta(y)+\ln\bigl(2\sqrt{\pi(u-1)}\lsp \bigr).
\end{equation}
The proof below gives the sharper bound
\begin{equation} \label{eq:cap-cost-gain}
0<K_u(\tau,\delta)\le uK_{\rm cov}(\tau,\delta)-\ln\frac{16}{\delta^3}.
\end{equation}
The finite cost $K_d$ is used for \thref{thm:large-floors}; the dimension-linear estimate in \eref{eq:cap-cost-gain} gives the simpler bounds used for \thref{thm:near-half}.

\begin{lemma}[Uniformization by Haar caps] \label{lem:common-uniform} Let $d\ge2$ and $L\ge1$ be integers, and let $\caM_{d,L}$ be the random measurement defined in the main text.
Let $c,\tau>0$ and $0<\delta\le1/10$. Put $\iota:=\sqrt{c}\lsp +\delta$. For every fixed admissible pair $(\ket{\psi_*},\ket{\nu_*})$,
\begin{equation} \label{eq:common-uniform}
\Pr\{\eta(\caM_{d,L})<c\}\le
\exp[K_d(\tau,\delta)]\Pr\{R_{\tau,\caM_{d,L}}(\psi_*,\nu_*)<\iota\}.
\end{equation}
\end{lemma}
\begin{lemma}[Beta distribution and Beta--Laplace bound] \label{lem:beta-laplace}
For a fixed $(\ket\psi,\ket\nu)\in\caS_d$ and a Haar orthonormal basis $\caB=\{\ket{\varphi_k}\}_{k=0}^{d-1}$ of $\caH$,
\begin{equation} \label{eq:beta-laplace-law}
g_{\caB}\sim\Beta \left(\frac{d-1}{2},\frac{d-1}{2}\right).
\end{equation}
For $t>0$, its Laplace transform satisfies
\begin{equation} \label{eq:beta-laplace-bound}
\Phi_d(t):=\E \rme^{-tg_{\caB}}
=\rme^{-t/2}\sum_{j=0}^\infty\frac{\left(t^2/16\right)^j}{j!(d/2)_j}
\le\exp\left(-\frac{t}{2}+\frac{t^2}{8d}\right).
\end{equation}
\end{lemma}
For $\tau,t>0$, set
\begin{equation} \label{eq:beta-correction}
\Delta_d(\tau,t):=\frac{t\tau}{2(1+\tau)}
\left(1+\frac{t}{(d+1)(1+\tau)}\right)^{-3/2}\le\frac{t\tau}{2(1+\tau)}\le\frac{t\tau}{2}.
\end{equation}
\begin{lemma}[Beta--Laplace fixed-pair bound] \label{lem:common-beta} Let $d\ge2$ and $L\ge1$ be integers, and let $\caM_{d,L}$ be the random measurement defined in the main text.
Let $\tau,t,\iota>0$. For every fixed admissible pair,
\begin{equation} \label{eq:fixed-beta-bound}
\Pr\{R_{\tau,\caM_{d,L}}<\iota\}\le
\left\{\rme^{t\iota^2}\bigl[\Phi_d(t)+\Delta_d(\tau,t)\bigr]\right\}^L.
\end{equation}
\end{lemma}
\begin{lemma}[Gaussian--QR fixed-pair bound] \label{lem:common-qr} Let $d\ge2$ and $L\ge1$ be integers, and let $\caM_{d,L}$ be the random measurement defined in the main text.
Let $\tau,s_*,\sigma,\iota>0$, where $s_*$ is a deterministic comparison scale. Define
\begin{equation} \label{eq:refined-qr-threshold}
x_{\QR}:=2\left(1+\frac{\tau}{s_*^2}\right)\left(s_*^2\iota^2+\sigma^2\right),
\end{equation}
and assume $x_{\QR}<1$. For every fixed admissible pair,
\begin{equation} \label{eq:refined-qr-bound}
\Pr\{R_{\tau,\caM_{d,L}}<\iota\}\le
\frac{4}{F_{L(d-1)}\left(L\min(d s_*^2,d-1)\right)F_L\left(Ld\sigma^2\right)}\exp\left[-\frac{Ld}{2}\Lambda(x_{\QR})\right].
\end{equation}
If also $s_*^2\ge(d-1)/d$ and $d\sigma^2\ge1$, the prefactor can be replaced by $4$:
\begin{equation} \label{eq:refined-qr-bound-2}
\Pr\{R_{\tau,\caM_{d,L}}<\iota\}\le4\exp\left[-\frac{Ld}{2}\Lambda(x_{\QR})\right].
\end{equation}
\end{lemma}

\begin{proposition}[Global Fisher-floor probability bounds] \label{prop:common-global} Let $d\ge2$ and $L\ge1$ be integers, and let $\caM_{d,L}$ be the random measurement defined in the main text.
Let $c,\tau>0$, $0<\delta\le1/10$, and $\iota:=\sqrt{c}\lsp +\delta$. For $t>0$,
\begin{equation} \label{eq:global-beta-cert}
\Pr\{\eta(\caM_{d,L})<c\}\le
\rme^{K_d(\tau,\delta)+Lt\iota^2}\bigl[\Phi_d(t)+\Delta_d(\tau,t)\bigr]^L.
\end{equation}
For $s_*,\sigma>0$ with $x_{\QR}<1$,
\begin{equation} \label{eq:global-qr-cert}
\Pr\{\eta(\caM_{d,L})<c\}\le
\frac{4\exp\left[K_d(\tau,\delta)-\frac{Ld}{2}\Lambda(x_{\QR})\right]}{F_{L(d-1)}\left(L\min(d s_*^2,d-1)\right)F_L\left(Ld\sigma^2\right)}.
\end{equation}
Under the additional hypotheses of \eref{eq:refined-qr-bound-2},
\begin{equation} \label{eq:global-simplified-qr}
\Pr\{\eta(\caM_{d,L})<c\}\le\frac{\delta^3}{4}\exp\left\{-d\left[\frac{L}{2}\Lambda(x_{\QR})-K_{\rm cov}(\tau,\delta)\right]\right\}.
\end{equation}
\end{proposition}

\begin{proof}[Proof of \lref{lem:common-trunc}]
We first establish the ordering, common zero set, and continuity. The ordering and the equality of zero sets follow termwise from \eref{eq:scalar-trunc}, including when $a=0$. Identify $\bbC$ with $\bbR^2$ and write the numerator as the real dot product of $a$ and $v$. In the region $|a|\ge\sqrt{\tau}\lsp|v|$, the normalized vector $a/|a|$ has derivative of operator norm $1/|a|$, so the gradient in $a$ has norm at most $|v|/|a|\le\tau^{-1/2}$. The gradient in $v$ has norm at most one. In the other region, the same argument with $v/|v|$ gives the bounds $\tau^{-1/2}$ and $|a|/(\sqrt{\tau}\lsp |v|)\le1$, respectively. The formulas agree on the interface, and $|h_\tau|\le\min\left\{|v|,|a|/\sqrt{\tau}\lsp \right\}$ gives continuity at zero inputs. Integrating along a line segment in each variable gives
\begin{equation} \label{eq:scalar-lipschitz}
|h_\tau(a,v)-h_\tau(a',v')|\le\tau^{-1/2}|a-a'|+|v-v'|.
\end{equation}
To pass from \eref{eq:scalar-lipschitz} to \eref{eq:resp-lipschitz}, apply the reverse triangle inequality to the vector of responses $h_\tau(a_{\ell k},v_{\ell k})/\sqrt{L}$. Minkowski then gives
\begin{equation}
|R_{\tau,\caM}(\psi,\nu)-R_{\tau,\caM}(\psi',\nu')|\le\frac{1}{\sqrt{\tau}\lsp }\left(\frac{1}{L}\sum_{\ell,k}|a_{\ell k}-a'_{\ell k}|^2\right)^{1/2}+\left(\frac{1}{L}\sum_{\ell,k}|v_{\ell k}-v'_{\ell k}|^2\right)^{1/2}.
\end{equation}
Parseval identifies the two sums with $\|\ket\psi-\ket{\psi'}\|^2$ and $\|\ket\nu-\ket{\nu'}\|^2$, respectively, proving \eref{eq:resp-lipschitz}.
\end{proof}

\begin{proof}[Proof of \lref{lem:common-uniform}]
We first record a spherical-cap bound for the tangent ket. For a unit ket $\ket\chi\in\bbC^m$ and $0<y<1$, let $A_m(y)$ be the normalized surface measure of the unit kets within Euclidean distance $y$ of $\ket\chi$. For every integer $m\ge1$,
\begin{equation} \label{eq:sphere-cap}
A_m(y)\ge\frac{\Gamma(m)}{2\sqrt{\pi}\lsp \Gamma(m+1/2)}\zeta(y)^{2m-1}\ge\frac{\zeta(y)^{2m-1}}{2\sqrt{\pi m}\lsp }.
\end{equation}
Orthogonal projection onto the real hyperplane perpendicular to $\ket\chi$ maps this cap onto a ball of radius $\zeta(y)$, with inverse surface-area Jacobian at least one. The ball-to-sphere volume ratio gives the first inequality; log-convexity~\cite[Sec.~5.5(iv)]{NISTDLMF}, $\Gamma(m+1/2)^2\le\Gamma(m)\Gamma(m+1)$, gives the second.

Fix a measurement realization with $\eta(\caM)<c$. \Eref{eq:dir-fisher-resp} and \lref{lem:common-trunc} supply an admissible pair with $R_{\tau,\caM}(\psi,\nu)<\sqrt{c}$. Draw a Haar orthonormal pair $(\ket{\psi'},\ket{\nu'})$. The response is unchanged when both kets are multiplied by the same phase. We may therefore align the phase of $\ket{\psi'}$ with $\ket\psi$; conditional on this aligned state, the tangent ket remains Haar in $\ket{\psi'}^\perp$. The projective cap
\begin{equation} \label{eq:projective-cap}
1-|\langle\psi|\psi'\rangle|^2\le\zeta(x)^2
\end{equation}
has probability $\zeta(x)^{2d-2}$, since the squared overlap is distributed as $\Beta(1,d-1)$. After phase alignment, it implies $\|\ket{\psi'}-\ket\psi\|\le x$. The Beta distribution follows by expressing a Haar ket as a normalized complex Gaussian vector: its first squared modulus is an exponential and the sum of the others is an independent $\Gamma(d-1,1)$ variable.

For a state in this cap, define the normalized projection
\begin{equation}
\ket{\widehat\nu}:=\frac{(\bbone-\proj{\psi'})\ket\nu}{\|(\bbone-\proj{\psi'})\ket\nu\|}.
\end{equation}
Orthogonality gives $|\langle\psi'|\nu\rangle|^2\le\zeta(x)^2<1$, so this ket is defined and
\begin{equation} \label{eq:cap-tangent-distance}
\|\ket{\widehat\nu}-\ket\nu\|^2=2\bigl(1-\sqrt{1-|\langle\psi'|\nu\rangle|^2}\lsp \bigr)\le x^2.
\end{equation}
Restrict $\ket{\nu'}$ to the spherical cap of radius $y$ about $\ket{\widehat\nu}$. Then $\|\ket{\nu'}-\ket\nu\|\le y+x$, and \eref{eq:resp-lipschitz} gives
\begin{equation}
R_{\tau,\caM}(\psi',\nu')\le R_{\tau,\caM}(\psi,\nu)+\frac{x}{\sqrt{\tau}\lsp }+y+x<\sqrt{c}\lsp +\delta=\iota.
\end{equation}
The projective-cap probability and \eref{eq:sphere-cap} in complex dimension $d-1$ therefore give a set of pairs of Haar measure at least
\begin{equation}
v_d:=\frac{\zeta(x)^{2d-2}\zeta(y)^{2d-3}}{2\sqrt{\pi(d-1)}\lsp }=\exp[-K_d(\tau,\delta)].
\end{equation}
The common phase does not affect this probability because the sublevel event is invariant under that phase. Thus, with $\mu$ the Haar probability measure on $\caS_d$, every measurement realization satisfies
\begin{equation} \label{eq:cap-sublevel-volume}
\bbone_{\{\eta(\caM)<c\}}v_d\le\int_{\caS_d}\bbone_{\{R_{\tau,\caM}(\psi',\nu')<\iota\}}\,\rmd\mu(\psi',\nu').
\end{equation}
Taking expectation over the bases and applying Fubini proves \eref{eq:common-uniform}: unitary invariance makes the probability in the resulting integral independent of the admissible pair.

Finally, retain both cap-radius factors in the prefactor. Since $\zeta(x),\zeta(y)\le\delta/2$ and $u-1\le u$,
\begin{equation} \label{eq:cap-prefactor}
\exp[K_u(\tau,\delta)]
=2\sqrt{\pi(u-1)}\lsp \zeta(x)^2\zeta(y)^3[\zeta(x)\zeta(y)]^{-2u}
\le\frac{\sqrt{\pi u}\lsp \delta^5}{16}[\zeta(x)\zeta(y)]^{-2u}
\le\frac{\delta^3}{16}\exp[uK_{\rm cov}(\tau,\delta)].
\end{equation}
For the last inequality, put $z=u\delta^4$ and use $\sqrt{\pi z}\lsp \rme^{-z}\le\sqrt{\pi/(2\rme)}\lsp <1$. This follows by maximizing $z\rme^{-2z}$ at $z=1/2$, with $\pi<2\rme$. Positivity of $K_u$ follows directly from its definition. This proves \eref{eq:cap-cost-gain} as well as \eref{eq:common-uniform}.
\end{proof}

\begin{proof}[Proof of \lref{lem:beta-laplace}]
Fix $(\ket\psi,\ket\nu)\in\caS_d$ and write the Haar basis as $\caB=\{\ket{\varphi_k}\}_{k=0}^{d-1}$. Define the amplitude vectors
\begin{equation}
 \bm a:=(\langle\varphi_k|\psi\rangle)_{k=0}^{d-1},\quad
 \bm v:=(\langle\varphi_k|\nu\rangle)_{k=0}^{d-1}\in\bbC^d.
\end{equation}
They form a Haar two-frame. Conditional on $\bm a$, the vector $\bm v$ is uniform on the complex unit sphere in $\bm a^\perp$. Almost surely all $a_k$ are nonzero. Put $b_k=|a_k|$ and $w_k=\rme^{-\rmi\arg a_k}v_k$. Then $\bm b=(b_k)_k$ is a real unit vector, and $\bm w=(w_k)_k$ is conditionally uniform on the complex unit sphere in $\bm b^\perp$. The row phases cancel in \eref{eq:basis-resp}, giving $g_{\caB}=\sum_k(\Re w_k)^2$.

Let $\bm x,\bm y$ be independent standard real Gaussian vectors in the real orthogonal complement of $\bm b$ in $\bbR ^d$, which has dimension $d-1$. A normalized complex Gaussian represents the conditional distribution of $\bm w$ as
\begin{equation}
 \bm w\overset{\rm law}=\frac{\bm x+\rmi\bm y}{\sqrt{\|\bm x\|^2+\|\bm y\|^2}\lsp },\quad
 g_{\caB}\overset{\rm law}=\frac{\|\bm x\|^2}{\|\bm x\|^2+\|\bm y\|^2}.
\end{equation}
The squared norms are independent $\chi_{d-1}^2$ variables, proving \eref{eq:beta-laplace-law}. To prove \eref{eq:beta-laplace-bound}, set $y=2g_{\caB}-1$. Symmetry of the Beta density and its even moments give
\begin{equation}
 \E y^{2j+1}=0,\quad \E y^{2j}=\frac{(1/2)_j}{(d/2)_j},\quad j=0,1,\ldots.
\end{equation} Expanding the exponential and using $(2j)!=4^jj!(1/2)_j$ yields the series in \eref{eq:beta-laplace-bound}. Since $|y|\le1$, termwise expectation is justified by absolute convergence. The inequality $(d/2)_j\ge(d/2)^j$ proves the final bound.
\end{proof}

\begin{proof}[Proof of \lref{lem:common-beta}]
Fix $(\ket\psi,\ket\nu)\in\caS_d$. We prove \eref{eq:fixed-beta-bound} by bounding the loss caused by denominator truncation. For one Haar basis $\caB=\{\ket{\varphi_k}\}_{k=0}^{d-1}$, use the amplitudes $a_k,v_k$ from \eref{eq:amplitudes} and define the two row contributions:
\begin{equation} \label{eq:row-contributions}
 u_k:=h_\tau^2(a_k,v_k),\quad
 w_k:=\begin{cases}\frac{[\Re (\overline{a}_kv_k)]^2}{|a_k|^2}&a_k\ne0,\\0&a_k=0.\end{cases}
\end{equation}
Thus $\sum_k u_k=R_{\tau,\caB}^2$, $\sum_k w_k=g_{\caB}$, and $0\le u_k\le w_k$. Telescoping the products of the factors $\rme^{-tu_k}$ and $\rme^{-tw_k}$, all in $[0,1]$, gives
\begin{equation} \label{eq:rowwise-laplace-loss}
0\le\rme^{-t\sum_ku_k}-\rme^{-t\sum_kw_k}
\le\sum_k\left(\rme^{-tu_k}-\rme^{-tw_k}\right).
\end{equation}
This estimate uses only row marginals. By \eref{eq:haar-row-laws}, $u_k$ is distributed as $y/(1+\tau)$ for $y\sim\Beta(1/2,d-1/2)$. Since $u_k$ tends to $w_k$ pointwise as $\tau$ decreases to zero, $w_k$ is distributed as $y$. Size-biasing the Beta density yields
\begin{equation} \label{eq:size-bias-identity}\E \left[y\rme^{-sy}\right]=\frac{1}{2d}\Psi_d(s).\end{equation}
Here $\Psi_d(s)=\E \rme^{-sz}$ and $z\sim\Beta (3/2,d-1/2)$. If $g\sim\Gamma(d+1,1)$ is independent of $z$, then $gz\sim\Gamma(3/2,1)$. Since $\E  g=d+1$ and the exponential is convex, Jensen conditional on $z$ gives
\begin{equation} \label{eq:beta-size-biased-bound}
\Psi_d(s)\le\E \rme^{-sgz/(d+1)}
=\left(1+\frac{s}{d+1}\right)^{-3/2}.
\end{equation}
We now apply \eref{eq:beta-size-biased-bound} to the row loss in \eref{eq:rowwise-laplace-loss}. The two marginal distributions just established give
\begin{align}
 \E \left(\rme^{-tu_k}-\rme^{-tw_k}\right)
 &=\E \left(\rme^{-ty/(1+\tau)}-\rme^{-ty}\right)
 =\frac{1}{2d}\int_{\frac{t}{1+\tau}}^t\Psi_d(s)\,\rmd s\notag\\
 &\le\frac{t\tau}{2d(1+\tau)}\left(1+\frac{t}{(d+1)(1+\tau)}\right)^{-3/2}
 =\frac{\Delta_d(\tau,t)}{d}. \label{eq:row-loss-integral}
\end{align}
The inequality uses monotonicity of $\Psi_d$ and \eref{eq:beta-size-biased-bound}. Summing \eref{eq:row-loss-integral} over the $d$ rows and applying \eref{eq:rowwise-laplace-loss} yields
\begin{equation} \label{eq:one-basis-laplace}
 \E \rme^{-tR_{\tau,\caB}^2}\le\Phi_d(t)+\Delta_d(\tau,t).
\end{equation}
Finally, the $L$ bases are independent. Exponential Markov gives
\begin{equation}
 \Pr\{R_{\tau,\caM_{d,L}}<\iota\}
 \le\rme^{Lt\iota^2}\prod_{\ell=0}^{L-1}\E \rme^{-tR_{\tau,\caB_\ell}^2}
 \le\left\{\rme^{t\iota^2}\bigl[\Phi_d(t)+\Delta_d(\tau,t)\bigr]\right\}^L,
\end{equation}
which is \eref{eq:fixed-beta-bound}. Replacing $\Psi_d(s)$ by its elementary bound $1$ in \eref{eq:row-loss-integral} gives the simpler correction in \eref{eq:beta-correction}. The size-biased estimate is retained to obtain the basis counts in \thref{thm:large-floors}.
\end{proof}

\begin{proof}[Proof of \lref{lem:common-qr}]
We prove \eref{eq:refined-qr-bound} by coupling the Haar amplitudes to Gaussian matrices. Fix $(\ket\psi,\ket\nu)\in\caS_d$. For each basis $\caB_\ell=\{\ket{\varphi_{\ell k}}\}_{k=0}^{d-1}$, define
\begin{equation}
 a_{\ell k}:=\langle\varphi_{\ell k}|\psi\rangle,\quad
 v_{\ell k}:=\langle\varphi_{\ell k}|\nu\rangle,\quad
 \bm a_\ell:=(a_{\ell k})_{k=0}^{d-1},\quad \bm v_\ell:=(v_{\ell k})_{k=0}^{d-1},\quad
 Q_\ell:=[\bm a_\ell\ \bm v_\ell].
\end{equation}
The $Q_\ell$ are independent Haar two-frames in $\bbC^d$. Independently adjoin random coefficients $r_\ell,s_\ell,z_\ell$, $\ell=0,\ldots,L-1$, with the mutually independent distributions in \eref{eq:gauss-qr}. These are distinct from the deterministic scale $s_*$. Set
\begin{equation} \label{eq:qr-coupled-matrices}
 Z_\ell:=\sqrt{d}\lsp Q_\ell\begin{pmatrix}r_\ell&z_\ell\\0&s_\ell\end{pmatrix}.
\end{equation}
\Lref{lem:gauss-qr} identifies the joint distribution of each frame and its triangular coefficients with that of a Gaussian QR decomposition. Therefore the matrices $Z_\ell$ have independent standard complex Gaussian entries, also independently across $\ell$. The $k$th row of $Z_\ell/\sqrt{d}\lsp $ is precisely $(r_\ell a_{\ell k},s_\ell v_{\ell k}+z_\ell a_{\ell k})$. Put $\tau_{\rm G}=\tau/s_*^2$ and $\bm z=(z_0,\ldots,z_{L-1})\in\bbC ^L$.

\emph{Gaussian coupling}---Define
\begin{equation} \label{eq:gauss-coupled-resp}
T(\bm z)^2:=\frac{1}{L}\sum_{\ell,k}h_{\tau_{\rm G}}^2(r_\ell a_{\ell k},s_\ell v_{\ell k}+z_\ell a_{\ell k}).
\end{equation}
Applying \eqsref{eq:adaptive-trunc}{eq:scalar-gauss-laws} to these Gaussian rows gives
\begin{equation} \label{eq:coupled-resp-law}
2(1+\tau_{\rm G})Ld\,T(\bm z)^2\sim\chi_{Ld}^2.
\end{equation}

\emph{Comparison costs}---Split the original response into its two denominator regions:
\begin{equation} \label{eq:qr-resp-split}
\scrI_\ell:=\left\{k:|a_{\ell k}|\ge\sqrt{\tau}\lsp |v_{\ell k}|\right\},\quad
A_\ell:=\sum_{k\in\scrI_\ell}h_\tau^2(a_{\ell k},v_{\ell k}),\quad
B_\ell:=\sum_{k\notin\scrI_\ell}h_\tau^2(a_{\ell k},v_{\ell k}).
\end{equation}
On $\scrI_\ell$, the new denominator is at least $r_\ell|a_{\ell k}|$, giving a bound $s_\ell^2h_\tau^2(a_{\ell k},v_{\ell k})$ for the unperturbed summand in $T(\bm0)^2$. On its complement, use the lower bound $\sqrt{\tau_{\rm G}}\lsp s_\ell|v_{\ell k}|$ instead, obtaining $s_*^2r_\ell^2h_\tau^2(a_{\ell k},v_{\ell k})$. Zero numerators cause no exception. Consequently
\begin{equation} \label{eq:qr-compare-cost}
T(\bm0)^2\le c_{\rm rad}:=\frac{1}{L}\sum_\ell\left(A_\ell s_\ell^2+B_\ell s_*^2r_\ell^2\right),\quad
c_{\rm par}:=\frac{1}{L}\sum_\ell|z_\ell|^2.
\end{equation}
By \lref{lem:sym-pert}, the mean squared response of each row over the two perturbation signs is at most its unperturbed value plus $|z_\ell a_{\ell k}|^2$. In each basis these terms sum to $|z_\ell|^2$, so
\begin{equation} \label{eq:off-diagonal-symmetry}
\frac{T(\bm z)^2+T(-\bm z)^2}{2}\le c_{\rm rad}+c_{\rm par}.
\end{equation}

\emph{Transfer to the Gaussian tail}---Conditional on $Q:=(Q_0,\ldots,Q_{L-1})$, the weights $A_\ell,B_\ell$ are fixed. The three independent events
\begin{equation} \label{eq:qr-three-events}
\sum_\ell A_\ell s_\ell^2\le s_*^2\sum_\ell A_\ell,\quad
\sum_\ell B_\ell r_\ell^2\le\sum_\ell B_\ell,\quad
c_{\rm par}\le\sigma^2
\end{equation}
have probabilities at least $F_{L(d-1)}\left(L\min\left(d s_*^2,d-1\right)\right)$, $1/2$, and $F_L\left(Ld\sigma^2\right)$, respectively. For the first, apply \lref{lem:weighted-gamma-mean} to $ds_\ell^2\sim\Gamma(d-1,1)$ with weights $A_\ell$, mean $(d-1)\sum_\ell A_\ell$, and threshold $d s_*^2\sum_\ell A_\ell$. Apply its at-mean bound to $dr_\ell^2\sim\Gamma(d,1)$ with weights $B_\ell$ for the second. A zero total weight makes the corresponding event certain. The third follows from $dL c_{\rm par}\sim\Gamma(L,1)$. Hence the sign-invariant event
\begin{equation} \label{eq:qr-general-transfer}
\caA:=\left\{c_{\rm rad}+c_{\rm par}\le s_*^2R_{\tau,\caM_{d,L}}^2+\sigma^2\right\}
,\quad
\Pr(\caA\mid Q)\ge p:=\frac{F_{L(d-1)}\left(L\min(d s_*^2,d-1)\right)F_L\left(Ld\sigma^2\right)}{2}.
\end{equation}
By \eref{eq:off-diagonal-symmetry}, on $\caA$ at least one of $T(\bm z)^2$ and $T(-\bm z)^2$ is at most $s_*^2R_{\tau,\caM_{d,L}}^2+\sigma^2$. The conditional distribution is invariant under $\bm z\mapsto-\bm z$, so the first alternative occurs together with $\caA$ with conditional probability at least $p/2$. On draws of $Q$ with $R_{\tau,\caM_{d,L}}<\iota$, this conditional bound applies to $T(\bm z)^2<s_*^2\iota^2+\sigma^2$. Averaging over $Q$ and using \eqsref{eq:coupled-resp-law}{eq:chi-square-tail} gives
\begin{equation} \label{eq:qr-transfer-to-chi}
\Pr\{R_{\tau,\caM_{d,L}}<\iota\}\le\frac{2}{p}\Pr\left\{T(\bm z)^2<s_*^2\iota^2+\sigma^2\right\}
\le\frac{2}{p}\exp\left[-\frac{Ld}{2}\Lambda(x_{\QR})\right].
\end{equation}
This proves \eref{eq:refined-qr-bound}.

Under the additional hypotheses, treat $c_{\rm rad}+c_{\rm par}$ as one weighted sum of independent Gamma variables of integer shapes $d-1,d,1$. Its conditional mean satisfies
\begin{equation} \label{eq:qr-cost-conditional}
\E (c_{\rm rad}+c_{\rm par}\mid Q)
=\frac{1}{L}\sum_\ell\left(\frac{d-1}{d}A_\ell+s_*^2B_\ell\right)+\frac{1}{d}
\le s_*^2R_{\tau,\caM_{d,L}}^2+\sigma^2.
\end{equation}
\Lref{lem:weighted-gamma-mean} gives $\Pr(\caA\mid Q)\ge1/2$. Using $p=1/2$ in the same transfer proves \eref{eq:refined-qr-bound-2}.
\end{proof}

\begin{proof}[Proof of \pref{prop:common-global}]
To prove \eref{eq:global-beta-cert}, apply the uniformization bound in \eref{eq:common-uniform} to the fixed-pair estimate in \eref{eq:fixed-beta-bound}, with $\iota:=\sqrt{c}\lsp +\delta$. Applying the same bound to \eref{eq:refined-qr-bound} gives \eref{eq:global-qr-cert}. For \eref{eq:global-simplified-qr}, combine \eref{eq:refined-qr-bound-2} with \eref{eq:cap-cost-gain}; the prefactors give $4\delta^3/16=\delta^3/4$.
\end{proof}

\section{Proof of \texorpdfstring{\thref{thm:five-bases}}{Theorem 1}: five-basis sufficiency} \label{supp:five-bases}

We use a rank-two measurement diagnostic throughout. The small-value tail proves the bound for $5\le L\le80$, while concentration about its mean handles $L\ge81$. Both estimates follow from the same rank-two distribution and cap argument. Define
\begin{equation} \label{eq:rank-two-orbit}
\Xi_d:=\left\{\proj\psi-\proj\nu:(\ket\psi,\ket\nu)\in\caS_d\right\}.
\end{equation}
For a basis $\caB=\{\ket{\varphi_k}\}_{k=0}^{d-1}$ and a Hermitian operator $O$, write $\diag_{\caB}O$ for the vector of diagonal entries in $\caB$ and $\|\cdot\|_1$ for its vector $\ell_1$ norm. For the equal mixture $\caM$ of bases $\caB_0,\ldots,\caB_{L-1}$, define
\begin{equation} \label{eq:rank-two-diagnostic}
\Pi_{\caB}(O):=\frac{1}{2}\|\diag_{\caB}O\|_1,\quad
\Pi_{\caM}(O):=\frac{1}{L}\sum_{\ell=0}^{L-1}\Pi_{\caB_\ell}(O).
\end{equation}
These functions are seminorms on the real space of Hermitian operators. For $(\ket\psi,\ket\nu)\in\caS_d$, $O=O_{\psi,\nu}$ from \eref{eq:general-dir-resp} belongs to $\Xi_d$. On the span of $\ket\psi,\ket\nu$ its eigenvalues are $+1,-1$, and it vanishes on the orthogonal complement. Cauchy--Schwarz, first over outcomes and then over bases, gives
\begin{equation} \label{eq:rank-two-fisher}
\Pi_{\caB}(O)=\sum_k|\Re(\overline{a}_kv_k)|\le\sqrt{g_{\caB}(\psi,\nu)},\quad
\eta(\caM)\ge\left[\min_{O\in\Xi_d}\Pi_{\caM}(O)\right]^2.
\end{equation}
The first inequality uses $\sum_k|a_k|^2=1$ and remains valid at zero amplitudes. Compactness of $\Xi_d$ and continuity of $\Pi_{\caM}$ ensure that the diagnostic minimum is attained. Thus a uniform lower bound on the measurement distance between orthogonal pure states yields a Fisher floor. Related measurement-norm estimates appear in Sec.~2 of Ref.~\cite{Voroninski2013} and in work on distinguishability by fixed measurements~\cite{MatthewsWehnerWinter2009,Sen2006,KuengZhuGross2016}.

\begin{lemma}[Rank-two distribution for Haar bases] \label{lem:rank-two-law}
Let $d\ge2$ and $L\ge1$ be integers, fix $O\in\Xi_d$, and put $n=d-1$. For a Haar orthonormal basis $\caB$,
\begin{equation} \label{eq:rank-two-beta}
\Pi_{\caB}(O)\sim\Beta(n,n).
\end{equation}
For $L$ independent Haar bases, the corresponding equal mixture $\caM_{d,L}$ satisfies, for $\iota>0$ and $t>0$,
\begin{align}
\Pr\{\Pi_{\caM_{d,L}}(O)<\iota\}&\le\frac{2^{-L/2}(4\iota)^{Ln}}{\sqrt{2\pi Ln}\lsp }, \label{eq:rank-two-fixed-tail}\\
\Pr\{\Pi_{\caM_{d,L}}(O)<\frac{1}{2}-t\}&\le\rme^{-2L(2n+1)t^2}. \label{eq:rank-two-gaussian-tail}
\end{align}
\end{lemma}

\begin{lemma}[Uniform rank-two bound] \label{lem:rank-two-uniform}
Let $d\ge2$ and $L\ge1$ be integers, put $n=d-1$, and fix $O\in\Xi_d$. Choose $r>0$, $b_0>\sqrt{2/\pi}$, and $0<\upsilon_*<1/2$ such that
\begin{equation} \label{eq:rank-two-params}
p_*:=1-\frac{\sqrt{2/\pi}\lsp }{b_0}-\frac{2\upsilon_*}{b_0}>0,\quad
\lambda_*:=\sqrt{1-4\upsilon_*^2}.
\end{equation}
Then
\begin{equation} \label{eq:rank-two-uniform-tail}
\Pr\left\{\eta(\caM_{d,L})<r^2\right\}\le\frac{4n\upsilon_*^{2-4n}}{p_*}\Pr\left\{\Pi_{\caM_{d,L}}(O)<\frac{r+b_0\upsilon_*}{\lambda_*}\right\}.
\end{equation}
\end{lemma}

\begin{proof}[Proof of \lref{lem:rank-two-law}]
Choose an orthonormal coordinate basis $\caB_{\stdState}$. In these coordinates, let $\bm g,\bm h\in\bbC^d$ be independent standard complex Gaussian vectors. Consider $H:=\bm g\bm g^\dag-\bm h\bm h^\dag$ conditional on the squared-norm difference being zero. We use the conditional distribution obtained by evaluating the joint densities at zero and dividing by the positive density of the squared-norm difference there.

Before conditioning, each diagonal entry has density $\rme^{-|x|}/2$. After conditioning, the diagonal density is proportional to $\exp(-\sum_k|x_k|)$ on $\sum_kx_k=0$. The volume of its $\ell_1$ ball of radius $2r$ scales as $r^n$, so
\begin{equation} \label{eq:rank-two-radial-laws}
R_0:=\frac{1}{2}\|\diag_{\caB_{\stdState}} H\|_1\sim\Gamma\left(n,\frac{1}{2}\right).
\end{equation}
The common squared norm $r_0:=\|\bm g\|^2=\|\bm h\|^2$ has density proportional to $r_0^{2d-2}\rme^{-2r_0}$, hence distribution $\Gamma(2n+1,1/2)$. The normalized kets $\ket\chi,\ket\xi$ are independent Haar kets, independent of $r_0$. Since $1-|\langle\chi|\xi\rangle|^2\sim\Beta(n,1)$, the positive eigenvalue of $H$ satisfies
\begin{equation}
\lambda=r_0\sqrt{1-|\langle\chi|\xi\rangle|^2}\lsp\sim\Gamma\left(2n,\frac{1}{2}\right)
\end{equation}
by the beta--gamma product identity~\cite{SiegristBeta}. Write $H=\lambda Y$, with $Y\in\Xi_d$. Unitary invariance and uniqueness of the invariant probability measure on $\Xi_d$ imply that the conditional distribution of $Y$ does not depend on $\lambda$. Thus $R_0=\lambda\Pi_{\caB_{\stdState}}(Y)$ gives
\begin{equation}
\E\Pi_{\caB_{\stdState}}(Y)^m=\frac{\E R_0^m}{\E\lambda^m}=\frac{(n)_m}{(2n)_m},\quad m=0,1,\ldots.
\end{equation}
These moments determine the distribution on $[0,1]$ and identify it as $\Beta(n,n)$. Haar invariance proves \eref{eq:rank-two-beta}.

For independent $X_1,\ldots,X_L\sim\Beta(n,n)$, drop the density factors $(1-x_\ell)^{n-1}\le1$ and extend the integration domain to the positive simplex. The Dirichlet integral gives
\begin{equation}
\Pr\left\{\sum_{\ell=1}^LX_\ell<L\iota\right\}\le\frac{[\Gamma(2n)/\Gamma(n)]^L(L\iota)^{Ln}}{\Gamma(Ln+1)}\le\frac{2^{-L/2}(4\iota)^{Ln}}{\sqrt{2\pi Ln}\lsp }.
\end{equation}
The last step uses Robbins's Stirling bounds~\cite{Robbins1955}, which give
\begin{equation} \label{eq:rank-two-gamma-ratio}
\frac{\Gamma(2n)}{\Gamma(n)}\le\frac{(4n/\rme)^n}{\sqrt{2}\lsp }\exp\left(\frac{1}{24n}-\frac{1}{12n+1}\right)\le\frac{(4n/\rme)^n}{\sqrt{2}\lsp },\quad
\Gamma(Ln+1)\ge\sqrt{2\pi Ln}\lsp (Ln/\rme)^{Ln}.
\end{equation}
Finally, apply \eref{eq:beta-laplace-bound} with dimension parameter $2n+1$. Exponential Markov's inequality yields
\begin{equation}
\Pr\left\{L^{-1}\sum_{\ell=1}^LX_\ell<\frac{1}{2}-t\right\}\le\inf_{u>0}\exp\left[-Lut+\frac{Lu^2}{8(2n+1)}\right]=\rme^{-2L(2n+1)t^2}.
\end{equation}
\end{proof}

\begin{proof}[Proof of \lref{lem:rank-two-uniform}]
Fix $O=\proj\psi-\proj\nu\in\Xi_d$. Sample $\ket{\psi'},\ket{\nu'}$ independently from the projective caps $1-|\langle\psi|\psi'\rangle|^2\le\upsilon_*^2$ and $1-|\langle\nu|\nu'\rangle|^2\le\upsilon_*^2$. Each cap has Haar measure $\upsilon_*^{2n}$. After phase alignment, write
\begin{equation}
\ket{\psi'}=\sqrt{1-s}\lsp\ket\psi+\sqrt{s}\lsp\ket\chi,\quad
\ket{\nu'}=\sqrt{1-t}\lsp\ket\nu+\sqrt{t}\lsp\ket\xi,\quad
\E s=\E t=\frac{n\upsilon_*^2}{n+1}.
\end{equation}
Here $\ket\chi$ and $\ket\xi$ are independent uniform unit kets in $\ket\psi^\perp$ and $\ket\nu^\perp$, respectively. The ratios $s/\upsilon_*^2$ and $t/\upsilon_*^2$ are independent $\Beta(n,1)$ variables, independent of the two kets. Set $O'=\proj{\psi'}-\proj{\nu'}$.

For a uniform unit vector $\bm z\in\bbC^n$, the squared real coordinate $(\Re z_1)^2$ is distributed as $\Beta(1/2,n-1/2)$ by the real Gaussian representation of the sphere. Its square-root moment is $\gamma_n:=\E|\Re z_1|=\Gamma(n)/[\sqrt{\pi}\lsp\Gamma(n+1/2)]$. If $\bm z,\bm w$ are independent uniform unit vectors and $A,B\ge0$, unitary invariance and Jensen's inequality give
\begin{equation} \label{eq:joint-sphere-mean}
\E|\Re(Az_1-Bw_1)|=\gamma_n\E\|A\bm z-B\bm w\|\le\gamma_n\sqrt{A^2+B^2}.
\end{equation}
For a fixed basis $\caB=\{\ket{\varphi_j}\}_{j=0}^{d-1}$, put $p_j=|\langle\varphi_j|\psi\rangle|^2$ and $q_j=|\langle\varphi_j|\nu\rangle|^2$. The projections of $\ket{\varphi_j}$ onto $\ket\psi^\perp$ and $\ket\nu^\perp$ have norms $\sqrt{1-p_j}\lsp$ and $\sqrt{1-q_j}$, respectively. Conditional on $s,t$, the linear term from each projector is twice the real part of an amplitude product. Their difference, after including the factor $1/2$ in $\Pi_{\caB}$ and using \eref{eq:joint-sphere-mean}, contributes at most
\begin{equation} \label{eq:joint-linear-mean}
\gamma_n\sum_j\sqrt{s(1-s)p_j(1-p_j)+t(1-t)q_j(1-q_j)}\lsp\le\gamma_n\sqrt{n(s+t)}.
\end{equation}
The inequality follows from Cauchy--Schwarz and $\sum_jp_j^2,\sum_jq_j^2\ge1/d$. The two remaining terms are $s(\proj\chi-\proj\psi)$ and $-t(\proj\xi-\proj\nu)$. Each projector difference has diagnostic at most one, so their contribution is at most $s+t$. Log-convexity gives $\Gamma(n+1)^2\le(n+1/2)\Gamma(n+1/2)^2$, hence $\gamma_n^2\le(n+1/2)/(\pi n^2)$. Radial averaging and Jensen's inequality imply, for every fixed mixture,
\begin{equation} \label{eq:rank-two-cap-mean}
\E\Pi_{\caM}(O'-O)\le\gamma_n n\sqrt{\frac{2}{n+1}}\lsp\upsilon_*+2\upsilon_*^2\le\sqrt{\frac{2}{\pi}}\lsp\upsilon_*+2\upsilon_*^2.
\end{equation}

If $\eta(\caM)<r^2$, \eref{eq:rank-two-fisher} supplies $O$ with $\Pi_{\caM}(O)<r$. By Markov's inequality, a fraction at least $p_*$ of its cap pairs satisfy $\Pi_{\caM}(O')<r+b_0\upsilon_*$. Every cap pair also obeys
\begin{equation}
|\langle\psi'|\nu'\rangle|\le\|\proj{\psi'}-\proj\psi\|_{\op}+\|\proj{\nu'}-\proj\nu\|_{\op}\le2\upsilon_*.
\end{equation}
Under unrestricted independent Haar sampling of $\ket{\psi'},\ket{\nu'}$, write $O'=\lambda Y$, where $\lambda=\sqrt{1-|\langle\psi'|\nu'\rangle|^2}$ is the positive eigenvalue and $Y\in\Xi_d$ is uniform and independent of $\lambda$ by unitary invariance. The squared overlap is distributed as $\Beta(1,n)$, so
\begin{equation}
\Pr\{\lambda\ge\lambda_*\}=1-\left(1-4\upsilon_*^2\right)^n\le4n\upsilon_*^2.
\end{equation}
For a measurement with $\eta(\caM)<r^2$, sampling a cap pair is equivalent to drawing two independent Haar kets conditioned on their respective caps around the witness. The product cap has Haar measure $\upsilon_*^{4n}$, and the conditional good-pair fraction is at least $p_*$. Thus the unrestricted event $\lambda\ge\lambda_*$ and $\Pi_{\caM}(Y)<(r+b_0\upsilon_*)/\lambda_*$ has probability at least $p_*\upsilon_*^{4n}$. Averaging over the measurement gives a lower bound of $p_*\upsilon_*^{4n}\Pr\left\{\eta(\caM_{d,L})<r^2\right\}$ for this joint probability. Under unrestricted sampling, independence of $\lambda$ and $Y$ separates the two events, and Haar invariance identifies the averaged diagnostic tail with the tail at any fixed $O\in\Xi_d$. This gives \eref{eq:rank-two-uniform-tail}.
\end{proof}

\subsection{Proof of the explicit floor and the five-basis refinement}
\emph{Remark (five-basis refinement).} A separate parameter choice in the same rank-two argument gives, for every integer $d\ge2$,
\begin{equation} \label{eq:five-basis-refinement}
\Pr\left\{\eta(\caM_{d,5})\ge10^{-8}\right\}\ge1-\rme^{-(d+23)/2}.
\end{equation}

\begin{proof}[Proof of \thref{thm:five-bases}]
Put $\eta=\eta(\caM_{d,L})$, $n=d-1$, $k=L-4$, and $r=\sqrt{c_L}\lsp=\rme^{-3L^{3/4}/k}/2$.

Suppose first that $5\le L\le80$. Choose $b_0=9/10$ and $\upsilon_*=3r/(kb_0)$ in \lref{lem:rank-two-uniform}. Combining it with \eref{eq:rank-two-fixed-tail} gives
\begin{equation} \label{eq:rank-two-QB}
\Pr\left\{\eta<r^2\right\}\le Q_*\sqrt{\frac{n}{2\pi L}}\lsp B_*^n,\quad
Q_*:=\frac{4\upsilon_*^2\,2^{-L/2}}{p_*},\quad
B_*:=\upsilon_*^{-4}\left[\frac{4(r+b_0\upsilon_*)}{\lambda_*}\right]^L.
\end{equation}
For $L=5,\ldots,11$, take rational upper bounds $u_L$ on $\upsilon_*$ with $10^4u_L=1,27,76,118,148,168,180$, respectively. Indeed, put $x_L=\lfloor 3000L^{3/4}/k\rfloor/1000$, evaluating the floor by comparing fourth powers. The exponential series gives $\rme^{3L^{3/4}/k}\ge\sum_{j=0}^{16}x_L^j/j!$, and substitution gives $\upsilon_*<u_L<1/20$. Using $\sqrt{2/\pi}\lsp<4/5$ gives $p_*>(1-20u_L)/9>0$. Squaring the coefficients in \eref{eq:rank-two-QB} then gives the rational comparisons
\begin{equation*}
 B_*^2\le\frac{[6(k+3)/5]^{2L}u_L^{2k}}{(1-4u_L^2)^L},\quad
 Q_*^2\le 2^{-L}\left(\frac{36u_L^2}{1-20u_L}\right)^2.
\end{equation*}
Substituting the seven values of $u_L$ yields
\begin{equation} \label{eq:rank-two-small-checks}
B_*<\frac{1}{2}\quad(5\le L\le11),\quad
Q_*<\begin{cases}\frac{1}{4000},&5\le L\le7,\\\frac{1}{2000},&8\le L\le11.\end{cases}
\end{equation}

For $12\le L\le80$, the function $\upsilon_*=(5/3k)\rme^{-3L^{3/4}/k}$ has its unique maximum between $L=14$ and $L=15$: the sign of its logarithmic derivative is the sign of $3L^{-1/4}(L+12)/(4k)-1$, whose first term decreases. Since $15^{3/4}>15/2$,
\begin{equation}
\upsilon_*<\frac{1}{6}\rme^{-45/22}<\frac{1}{45},\quad
p_*>\frac{5}{81},\quad
-L\ln\lambda_*\le\frac{2L\upsilon_*^2}{1-4\upsilon_*^2}<\frac{1}{12},\quad
Q_*<\frac{4}{125}\,2^{-L/2}\le\frac{1}{2000}.
\end{equation}
To bound $B_*$, regard $L$ as real and set
\begin{equation}
f(L):=\ln\left(B_*\lambda_*^L\right)=4\ln\frac{3k}{10}+(L+4)\ln2+L\ln\frac{k+3}{k}-3L^{3/4},\quad
f''(L)=\frac{9}{16L^{5/4}}-\frac{4k+9}{k(k+3)^2}.
\end{equation}
The ratio of the positive term in $f''$ to the magnitude of its negative term increases with $k>0$, since its logarithmic derivative is
\begin{equation*}
\frac{1}{k}-\frac{4}{4k+9}+\frac{2}{k+3}-\frac{5}{4(k+4)}>0.
\end{equation*}
As $f'(12)<0$, the function $f$ has no interior maximum on $[12,81]$. The endpoint bounds $f(12),f(81)<-9/10$ therefore give $B_*<\rme^{-49/60}<1/2$. Together with \eref{eq:rank-two-small-checks}, this proves $B_*<1/2$ and $Q_*/\sqrt{2\pi L}\lsp<1/14000$ throughout $5\le L\le80$. Using $\sqrt{n}\lsp2^{-n}\le\rme^{-n/2}$ and $\rme^{19/2}<14000$ in \eref{eq:rank-two-QB} yields the required bound $\rme^{-(n+19)/2}$.

Now let $L\ge81$, and set $s=L^{-1/4}$ and $t=[(\ln(2L)+1/4)/L]^{1/2}$. We have $s\le1/3$ and $t\le4s/5$: the function $16\sqrt{L}\lsp/25-\ln(2L)-1/4$ is increasing and positive at $81$. The function $1-8s/5-2s^4-\rme^{-3s}$ is concave on $[0,1/3]$ and nonnegative at both endpoints. Consequently,
\begin{equation}
r\le\frac{1}{2}\rme^{-3s}\le\frac{1}{2}-\frac{4s}{5}-s^4\le\frac{1}{2}-t-\frac{1}{L}.
\end{equation}
Apply \lref{lem:rank-two-uniform} with $\upsilon_*=1/(2L)$ and $b_0=1$. Then $p_*>1/8$ and $1-\lambda_*\le L^{-2}$, so $(r+\upsilon_*)/\lambda_*\le1/2-t$. \Eref{eq:rank-two-gaussian-tail} gives
\begin{equation}
\Pr\left\{\eta<r^2\right\}\le\frac{4n\upsilon_*^{2-4n}}{p_*}\rme^{-(4n+2)Lt^2}\le\frac{2n}{L^4}\rme^{-n-1/2}\le\rme^{-(n+19)/2}.
\end{equation}
The last inequality follows from $n\rme^{-n/2}\le2/\rme$ and $4\rme^8<81^4$. Finally,
\begin{equation*}
\frac{\rmd}{\rmd L}\ln c_L=\frac{3(L+12)}{2L^{1/4}(L-4)^2}>0,\quad c_L\longrightarrow\frac{1}{4}.
\end{equation*}
Direct substitution gives $c_5>4.8\times10^{-10}$, completing the proof.
\end{proof}

\begin{proof}[Proof of \eref{eq:five-basis-refinement}]
Use \lref{lem:rank-two-uniform} with $L=5$, $r=10^{-4}$, $b_0=13/16$, and $\upsilon_*=4r/b_0=4/8125$. In the notation of \eref{eq:rank-two-QB}, the inequalities $\sqrt{2/\pi}\lsp<4/5$ and $\upsilon_*<1/2000$ give
\begin{equation}
p_*>\frac{1}{80},\quad
B_*=\frac{5}{4}\left(\frac{13}{16}\right)^4\lambda_*^{-5}<\frac{11}{20},\quad
Q_*<\frac{1}{60000}.
\end{equation}
Here $\lambda_*^{-5}<1000/999$ suffices for the middle inequality. Thus, for $d\ge2$,
\begin{equation}
\Pr\left\{\eta(\caM_{d,5})<10^{-8}\right\}\le\frac{1}{60000}\sqrt{\frac{d-1}{10\pi}}\lsp\left(\frac{11}{20}\right)^{d-1}<\frac{\rme^{-(d-1)/2}}{200000}<\rme^{-(d+23)/2}.
\end{equation}
The last two steps use $11/20<\rme^{-7/12}$, $\sup_{x>0}\sqrt{x}\lsp\rme^{-x/12}=\sqrt{6/\rme}\lsp<3/2$, and $\rme^{12}<200000$.
\end{proof}

\section{Proof of \texorpdfstring{\thref{thm:large-floors}}{Theorem 2}: large constant Fisher floors} \label{supp:large-floors}

The proof has two ranges: the Beta--Laplace bound handles finitely many low dimensions, and a derivative bound continues the Gaussian--QR estimate from one endpoint. All parameters below are explicit rational numbers. The phase reduction in \lref{lem:common-uniform} lowers the cap cost and helps control the failure probability at small $d$.

For $x,y>0$, define
\begin{equation} \label{eq:gamma-log-lower}
\Theta(x,y):=-y+x\left(1+\ln\frac{y}{x}\right)-\frac{1}{2}\ln\frac{44x}{7}-\frac{1}{12x}.
\end{equation}
For every integer $m\ge1$, \eref{eq:gamma-bounds} and Robbins's upper Stirling bound~\cite{Robbins1955} imply $\ln F_m(y)\ge\Theta(m,y)$. Explicitly, the bounds used are
\begin{equation}
m!\le\sqrt{2\pi m}\lsp \left(\frac{m}{\rme}\right)^m\rme^{1/(12m)},\quad 2\pi<\frac{44}{7}.
\end{equation} The smooth extension $\Theta(x,y)$ is used only for the continuation calculation below.

\begin{lemma}[Continuation of the QR bound] \label{lem:large-cont} Let $d\ge2$ and $L\ge1$ be integers, and let $\caM_{d,L}$ be the random measurement defined in the main text.
Let $D\ge2$ be an integer. Fix $c,\tau,s_*,\sigma>0$ and $0<\delta\le1/10$. Use $\iota=\sqrt{c}\lsp+\delta$ and $x_{\QR}$ from \eref{eq:refined-qr-threshold}, and assume $x_{\QR}<1$ and $s_*^2\le(D-1)/D$. For real $u\ge D$, set
\begin{align}
\Upsilon_L^{\QR}(u)&:=\frac{Lu}{2}\Lambda(x_{\QR})-K_u(\tau,\delta)
+\Theta\left(L(u-1),Lu s_*^2\right)+\Theta\left(L,LD\sigma^2\right)-\ln4, \label{eq:qr-cont-E}\\
\mu_D&:=\frac{L}{2}\Lambda(x_{\QR})-K_{\rm cov}(\tau,\delta)-L\Lambda\left(s_*^2\right)-\frac{1}{2(D-1)}-\frac{1}{D}. \label{eq:qr-cont-slope}
\end{align}
For every integer $d\ge D$,
\begin{equation} \label{eq:qr-cont-bound}
\Pr\{\eta(\caM_{d,L})<c\}\le\rme^{-\Upsilon_L^{\QR}(d)},\quad
\Upsilon_L^{\QR}(d)\ge\Upsilon_L^{\QR}(D)+(d-D)\mu_D.
\end{equation}
\end{lemma}

\begin{table}[htb]
\caption{Parameters for \thref{thm:large-floors}, with baseline basis count $L_0$. The two tuples have orders $(\tau_d,\delta,t_d)$ and $\left(\tau,\delta,s_*^2,\sigma^2\right)$, respectively. In particular, the third QR entry is the squared scale $s_*^2$.}
 \label{tab:large-params}
\centering\setlength{\tabcolsep}{12pt}\renewcommand{\arraystretch}{1.5}
\begin{tabular}{ccccc}
\toprule
$\displaystyle L_0$ & $\displaystyle c$ & $\displaystyle D$ & $\displaystyle 2\le d<D$: Beta--Laplace & $\displaystyle d\ge D$: Gaussian--QR\\
\midrule
$\displaystyle 71$ & $\displaystyle \frac{1}{10}$ & $\displaystyle 13$ & $\displaystyle \left(\frac{2\cdot47^d}{25\cdot100^d},\,\frac{13}{500},\,3d-2\right)$ & $\displaystyle \left(\frac{1}{60},\,\frac{1}{40},\,\frac{7}{12},\,\frac{1}{80}\right)$\\
\addlinespace[6pt]
$\displaystyle 271$ & $\displaystyle \frac{1}{4}$ & $\displaystyle 23$ & $\displaystyle \left(\frac{3\cdot17^d}{200\cdot25^d},\,\frac{3}{250},\,\frac{21d}{20}\right)$ & $\displaystyle \left(\frac{1}{128},\,\frac{1}{80},\,\frac{11}{16},\,\frac{1}{64}\right)$\\
\bottomrule
\end{tabular}
\end{table}

For the parameters under consideration, denote the negative logarithm of the right-hand side of \eref{eq:global-beta-cert} by
\begin{equation} \label{eq:beta-exponent}
\Upsilon_{d,L}^{\Beta}:=-K_d(\tau,\delta)-L\left\{t\iota^2+\ln[\Phi_d(t)+\Delta_d(\tau,t)]\right\}.
\end{equation}
The positive series in \eref{eq:beta-laplace-bound} permits rigorous finite evaluation. If $s_j=\left(t^2/16\right)^j/[j!(d/2)_j]$ and $r_N=t^2/[16(N+1)(d/2+N)]<1$, its decreasing term ratios give
\begin{equation} \label{eq:beta-series-enclosure}
\rme^{-t/2}\sum_{j=0}^N s_j
\le\Phi_d(t)\le
\rme^{-t/2}\left(\sum_{j=0}^N s_j+\frac{s_{N+1}}{1-r_N}\right).
\end{equation}
The comparisons in \tref{tab:large-certs} can all be checked with rational arithmetic. Use $N=16$ in \eref{eq:beta-series-enclosure} and bound $\rme^{-t/2}$ above by the reciprocal of $\sum_{j=0}^{48}(t/2)^j/j!$. Enclose each square root to within $10^{-8}$ by comparing the squares of neighboring grid points. For logarithms, reduce the argument to $y\in[1,2]$ by inversion and scaling by powers of two. With $z=(y-1)/(y+1)$ and $S=2\sum_{j=0}^{11}z^{2j+1}/(2j+1)$, the positive series gives
\begin{equation*}
 S\le\ln y\le S+\frac{2z^{25}}{25(1-z^2)}.
\end{equation*}
Apply this bound also to $\ln 2$. For the cap cost, use $\pi<22/7$ and the increase of $\zeta$ with its radius. For the QR tail, use the decrease of $\Lambda$ on $(0,1)$. Substituting these one-sided bounds gives the strict inequalities in \tref{tab:large-certs}. We write $\mu_D(L)$ when the dependence on $L$ is needed.

\begin{table}[htb]
\caption{Certified strict lower bounds at $L=L_0$ for the parameters in \tref{tab:large-params}. Both QR thresholds satisfy $0<x_{\QR}<1$.}
 \label{tab:large-certs}
\centering\setlength{\tabcolsep}{12pt}\renewcommand{\arraystretch}{1.5}
\begin{tabular}{ccccc}
\toprule
$\displaystyle L_0$ & $\displaystyle D$ & $\displaystyle \min_{2\le d<D}\frac{\Upsilon_{d,L_0}^{\Beta}}{d+10}$ & $\displaystyle \Upsilon_{L_0}^{\QR}(D)$ & $\displaystyle \mu_D(L_0)$\\
\midrule
$\displaystyle 71$ & $\displaystyle 13$ & $\displaystyle \frac{6}{5}$ & $\displaystyle 28$ & $\displaystyle \frac{7}{2}$\\
\addlinespace[6pt]
$\displaystyle 271$ & $\displaystyle 23$ & $\displaystyle \frac{1001}{1000}$ & $\displaystyle 39$ & $\displaystyle \frac{5}{4}$\\
\bottomrule
\end{tabular}
\end{table}

\begin{proof}[Proof of \lref{lem:large-cont}]
For $d\ge D$, the assumption $s_*^2\le(D-1)/D\le(d-1)/d$ removes the minimum from the first Gamma CDF in \eref{eq:global-qr-cert}. Bound its logarithm by $\Theta\left(L(d-1),Ld s_*^2\right)$. For the second, monotonicity gives $F_L\left(Ld\sigma^2\right)\ge F_L\left(LD\sigma^2\right)$, whose logarithm is at least $\Theta\left(L,LD\sigma^2\right)$. This proves the probability estimate in \eref{eq:qr-cont-bound}. To prove its second inequality, differentiate \eref{eq:gamma-log-lower}:
\begin{equation} \label{eq:qr-cont-derivative}
\frac{\rmd }{\rmd u}\Theta\left(L(u-1),Lu s_*^2\right)
=-L\Lambda\left(s_*^2\right)+L\left(\ln\frac{u}{u-1}-\frac{1}{u}\right)-\frac{1}{2(u-1)}+\frac{1}{12L(u-1)^2}.
\end{equation}
The bracket is nonnegative because $\ln[u/(u-1)]=\int_{u-1}^u t^{-1}\,\rmd t\ge1/u$. The last term is positive, and $u\ge D$ bounds the negative term below by $-1/[2(D-1)]$. Also, \eref{eq:finite-cap-cost} gives
\begin{equation}
 \frac{\rmd }{\rmd u}K_u(\tau,\delta)
 =K_{\rm cov}(\tau,\delta)-\delta^4+\frac{1}{2(u-1)}
 \le K_{\rm cov}(\tau,\delta)+\frac{1}{D}.
\end{equation}
For $u\ge D\ge2$, $1/[2(u-1)]\le1/D$. Combining these bounds yields $\partial_u\Upsilon_L^{\QR}(u)\ge\mu_D$. Integration over $[D,d]$ proves the second inequality in \eref{eq:qr-cont-bound}.
\end{proof}

For $x\ge1$, $y>0$, and $a\ge1$, direct substitution gives
\begin{equation} \label{eq:theta-scaling}
\Theta(ax,ay)-a\Theta(x,y)=\frac{a-1}{2}\ln\frac{44x}{7}-\frac{1}{2}\ln a+\frac{a-a^{-1}}{12x}\ge0,
\end{equation}
because $\ln(44x/7)>1$ and $\ln a\le a-1$.

\begin{proof}[Proof of \thref{thm:large-floors}]
For each row of \tref{tab:large-params}, put $\gamma=71/60$ when $L_0=71$ and $\gamma=1$ when $L_0=271$. We prove failure probability at most $\exp[-\gamma L(d+10)/L_0]$.

At $L=L_0$, \tref{tab:large-certs} gives $\Upsilon_{d,L_0}^{\Beta}>\gamma(d+10)$ for $2\le d<D$. For $d\ge D$, the QR endpoint exceeds $\gamma(D+10)$ and the slope exceeds $\gamma$, so \lref{lem:large-cont} gives $\Upsilon_{L_0}^{\QR}(d)>\gamma(d+10)$.

For $L\ge L_0$, keep all other parameters fixed and set $a=L/L_0$. Equations~\eqref{eq:beta-exponent}, \eqref{eq:qr-cont-E}, and \eqref{eq:theta-scaling} imply
\begin{align}
\Upsilon_{d,L}^{\Beta}&=a\Upsilon_{d,L_0}^{\Beta}+(a-1)K_d\ge a\Upsilon_{d,L_0}^{\Beta}, \label{eq:beta-larger-L}\\
\Upsilon_L^{\QR}(d)&\ge a\Upsilon_{L_0}^{\QR}(d)+(a-1)(K_d+\ln4)\ge a\Upsilon_{L_0}^{\QR}(d). \label{eq:qr-larger-L}
\end{align}
The two arguments of $\Theta$ scale by $a$ in each occurrence, and their first arguments at $L_0$ are at least one. Positivity of $K_d$ completes both comparisons. Applying \eqref{eq:global-beta-cert} for $d<D$ and \eqref{eq:qr-cont-bound} for $d\ge D$ proves \eref{eq:large-floor-bounds}.
\end{proof}

\section{Proof of \texorpdfstring{\thref{thm:near-half}}{Theorem 3}: approximate universal Fisher symmetry} \label{supp:near-half}

To prove \eref{eq:near-half}, we choose $\delta=\varepsilon/24$ and apply \eqsref{eq:global-beta-cert}{eq:global-simplified-qr}. For $d\varepsilon<13$, we choose the truncation scale using the exponential upper bound on the Beta--Laplace transform. For $d\varepsilon\ge13$, the choice $\sigma^2=\varepsilon/13$ meets the condition $d\sigma^2\ge1$ for the simplified QR bound. The remaining estimates are elementary.
\begin{proof}[Proof of \thref{thm:near-half}]
Set $\delta=\varepsilon/24$, $c=1/2-\varepsilon$, and $\iota=\sqrt{c}\lsp +\delta$. By hypothesis, $0<\varepsilon<1/2$. Put
\begin{equation}
u=\ln(1/\varepsilon),\quad w=L\varepsilon^2,
\quad w_0=12(1+u),\quad h=\ln\frac{16}{\delta^3},\quad \gamma=\frac{1}{24}.
\end{equation}
Then $u>\ln2$ and $w\ge w_0$. For fixed $d,\varepsilon$ and either parameter choice below, the negative logarithms supplied by \eqsref{eq:global-beta-cert}{eq:global-simplified-qr}, with the cap bound \eref{eq:cap-cost-gain}, are affine functions of $w$ with negative constant terms. Indeed, $dK_{\rm cov}-h\ge K_d>0$; the QR bound has the additional cost $\ln4$. Dividing either exponent by $w$ therefore gives an increasing function. It suffices to check the target rate at $w=w_0$. This is an evaluation of the scalar bound at a real argument, not an assumption that $w_0/\varepsilon^2$ is an integer basis count. The elementary tangent bound $\sqrt{1/2-\varepsilon}\lsp \le3/4-\varepsilon$ gives
\begin{equation} \label{eq:threshold}
\iota^2\le\frac{1}{2}-
\left(\frac{15}{16}+\frac{47\varepsilon}{576}\right)\varepsilon.
\end{equation}
Indeed, the difference of the squared sides of the tangent bound is $(\varepsilon-1/4)^2$.

Both truncation parameters below satisfy $0<\tau\le\varepsilon/32$. The cap cost in \eref{eq:covering-cost} gives
\begin{equation} \label{eq:cap-bounds}
h>12+3u,\quad
K_{\rm cov}(\tau,\delta)
\le\frac{96}{5}+5u+\ln\frac{\varepsilon}{32\tau}.
\end{equation}
To bound the constants, use $\sqrt{\tau}\lsp \le1/8$ and $x,y\le1/96$ in the cap-radius formulas. The constant part of $K_{\rm cov}$ is at most
\begin{equation}
21\ln2+4\ln3+2\ln\left(\frac{9}{8}\right)
-2\ln\left(1-192^{-2}\right)+48^{-4}<\frac{96}{5}.
\end{equation}
The elementary bounds $\ln2<52/75$, $\ln3<11/10$, $\ln(9/8)<19/160$, and $-2\ln\left(1-192^{-2}\right)+48^{-4}<1/400$ suffice. Also $h=13\ln2+3\ln3+3u>12+3u$, using $\ln2>9/13$ and $\ln3>1$.

\paragraph{Low dimensions: $d\varepsilon<13$.}
Set
\begin{equation} \label{eq:low-params}
\beta=\frac{3}{2}-\frac{9\varepsilon}{8},\quad
\tau=\frac{\varepsilon}{32}\rme^{-\beta d\varepsilon},\quad
t=3d\varepsilon.
\end{equation}
Since $\beta>15/16$, the stated range of $\tau$ holds. Equations~\eqref{eq:beta-laplace-bound} and \eqref{eq:beta-correction} apply, and the exponent in the former is exactly $-t/2+t^2/(8d)=-\beta d\varepsilon$, so
\begin{equation}
\Phi_d(t)+\Delta_d(\tau,t)
\le\rme^{-\beta d\varepsilon}
\left(1+\frac{3d\varepsilon^2}{64}\right).
\end{equation}
Together with \eqref{eq:threshold} and $\ln(1+x)\le x$, this gives fixed-pair rate $A(\varepsilon)d\varepsilon^2$ per basis, where
\begin{equation}
A(\varepsilon)=\frac{105}{64}+\frac{47\varepsilon}{192}.
\end{equation}
Equation~\eqref{eq:cap-bounds} and $d\varepsilon<13$ give
\begin{equation}
K_{\rm cov}\le\frac{387}{10}+5u-\frac{117\varepsilon}{8}.
\end{equation}
Define the scalar budget
\begin{equation}
g(u)=12(1+u)\left[A\left(\rme^{-u}\right)-\gamma\right] -\left(\frac{387}{10}+5u-\frac{117\rme^{-u}}8\right) =\frac{1135u-1561}{80} +\frac{281+47u}{16}\rme^{-u}.
\end{equation}
It satisfies
\begin{equation} \label{eq:low-budget}
g'(u)=\frac{227-(234+47u)\rme^{-u}}{16}>5,\quad
g(\ln2)=\frac{2505\ln2-1717}{160}>0.
\end{equation}
For the derivative, use $\rme^{-u}\le1/2$ and $u\rme^{-u}\le1/\rme<3/8$; for the value, use $\ln2>56/81$. At $w=w_0$, \eqsref{eq:global-beta-cert}{eq:cap-cost-gain} give a negative logarithm at least $\gamma dw_0+dg(u)+h$. Its excess over $\gamma(d+16)w_0$ satisfies
\begin{equation}
dg(u)+h-16\gamma w_0>2g(u)+4-5u>0.
\end{equation}
The last expression has derivative greater than $5$ and, at $u=\ln2$, exceeds $4-5\ln2>1/2$. This proves the baseline bound in the low-dimensional range.

\paragraph{High dimensions: $d\varepsilon\ge13$.}
Set
\begin{equation} \label{eq:high-params}
\tau=\frac{\varepsilon}{32},\quad s_*=1,\quad \sigma^2=\frac{\varepsilon}{13}.
\end{equation}
Then $s_*^2\ge(d-1)/d$ and $d\sigma^2\ge1$. Equation~\eqref{eq:threshold} gives
\begin{equation}
x_{\QR}\le\left(1+\frac{\varepsilon}{32}\right)
\left(1-\frac{179\varepsilon}{104}-\frac{47\varepsilon^2}{288}\right),
\quad
z:=1-x_{\QR}\ge\frac{703\varepsilon}{416}+\frac{\varepsilon^2}{5}.
\end{equation}
In particular, $0<x_{\QR}<1$. Retaining three terms of the logarithm series yields
\begin{equation} \label{eq:qr-rate}
\frac{1}{2}\Lambda(x_{\QR})
\ge\frac{z^2}{4}+\frac{z^3}{6}+\frac{z^4}{8}
\ge b(\varepsilon)\varepsilon^2,
\quad
b(\varepsilon)=\frac{57}{80}+\frac{97\varepsilon}{100}
+\frac{13\varepsilon^2}{10}.
\end{equation}
The coefficient comparisons in this estimate are recorded below. Using $\ln4/d<(7/65)\varepsilon$, define
\begin{align}
f(u)&=12(1+u)\left[b\left(\rme^{-u}\right)-\gamma\right]
-\left(\frac{96}{5}+5u+\frac{7\rme^{-u}}{65}\right)\notag\\
&=\frac{61u}{20}-\frac{223}{20}
+\rme^{-u}\left(\frac{3748}{325}+\frac{291u}{25}\right)
+\frac{78}{5}(1+u)\rme^{-2u}>\frac{3}{20}.
 \label{eq:high-budget}
\end{align}
The scalar inequality is proved below. Since $d\ge13\rme^u$ and $\rme^u\ge\rme u$,
\begin{equation}
df(u)+h>\frac{39}{20}\rme^u+12+3u>12+8u>8(1+u)=16\gamma w_0,
\end{equation}
where $39\rme/20>5$ follows from $\rme>8/3$. Thus the negative logarithm in \eref{eq:global-simplified-qr}, at $w=w_0$, exceeds $\gamma(d+16)w_0$. The initial scaling argument extends both dimensional ranges to all $w\ge w_0$, proving \eref{eq:near-half}.
\end{proof}

\paragraph{Elementary checks for the high-dimensional estimates.}
Put $a=703/416$. The coefficient comparisons in \eqref{eq:qr-rate} are
\begin{equation}
\frac{a^2}{4}>\frac{57}{80},\quad
\frac{a}{10}+\frac{a^3}{6}>\frac{97}{100},\quad
\frac{1}{100}+\frac{a^2}{10}+\frac{a^4}{8}>\frac{13}{10}.
\end{equation}
To prove \eref{eq:high-budget}, direct differentiation gives
\begin{equation}
f''(u)=\rme^{-u}\left[\frac{291}{25}(u-1)-\frac{7}{65}+\frac{312}{5}u\rme^{-u}\right]>0.
\end{equation}
For $\ln2\le u\le2$, the bracket exceeds $394/325$ by $u>2/3$ and $\rme^{-u}\ge\rme^{-2}>1/8$; for $u\ge2$, positivity is immediate. Choose $u_0=\ln12$, so the exponential terms become rational. The bounds
\begin{equation}
\frac{149}{60}<\ln12<\frac{87}{35}
\end{equation}
follow from $\ln12=4\ln2-\ln(4/3)$ and the positive series $\ln[(1+z)/(1-z)]=2\sum_{j\ge0}z^{2j+1}/(2j+1)$: retain three terms at $z=1/3$ and two at $z=1/7$, bounding each tail geometrically. Substitution gives
\begin{align}
f(u_0)&=-\frac{78629}{7800}+\frac{2477}{600}\ln12>\frac{1}{6},\\
0<f'(u_0)&=\frac{4603}{1560}-\frac{89}{75}\ln12<\frac{1}{200}.
\end{align}
Since $u-u_0\ge-\ln6>-2$, convexity yields
\begin{equation}
f(u)\ge f(u_0)+(u-u_0)f'(u_0)>\frac{1}{6}-\frac{2}{200}=\frac{47}{300}>\frac{3}{20}.
\end{equation}
This proves \eref{eq:high-budget} without numerical certification.

\section{Basic properties of Fisher floors} \label{supp:structural-proofs}
We establish three properties of a specified measurement: the mixing inequalities in \pref{prop:mean-monotone}, the upper bounds imposed by a fixed basis count in \pref{prop:basis-ceiling}, and the computable lower bound in \pref{prop:frame-cert}. These results do not require Haar-random bases. The mixing result also implies monotonicity of the expected floor for exchangeable measurement ensembles.

\subsection{Proof of \texorpdfstring{\pref{prop:mean-monotone}}{Proposition 6}: mixing and monotonicity of the expected floor} \label{supp:mixing}
\begin{proof}[Proof of \pref{prop:mean-monotone}]
Multiplying $E_{ji}$ by $q_j$ multiplies the corresponding term in \eref{eq:general-dir-resp} by $q_j$. This also holds when $q_j=0$ or the original outcome probability vanishes, since the contribution is then zero. Thus
\begin{equation} \label{eq:mixture-floor}
g_{\caM(q)}=\sum_{j=1}^m q_jg_{\caN_j},\quad
\eta(\caM(q))\ge\sum_{j=1}^m q_j\eta(\caN_j).
\end{equation}
Taking the infimum gives \eref{eq:floor-concavity}.

For $m\ge2$, let $\caM_{m-1}^{(-j)}$ be the uniform mixture of the $m-1$ POVMs obtained by omitting $\caN_j$. In the average over $j$, each component occurs $m-1$ times with total weight $1/m$. By \eref{eq:mixture-floor}, the average response is therefore $g_{\caM_m}$. Taking the infimum gives
\begin{equation} \label{eq:leave-one-out}
\eta(\caM_m)\ge\frac{1}{m}\sum_{j=1}^m\eta\bigl(\caM_{m-1}^{(-j)}\bigr).
\end{equation}
By exchangeability, each $\caM_{m-1}^{(-j)}$ has the same floor distribution as $\caM_{m-1}$. Taking expectations proves the result.
\end{proof}

\subsection{Fisher-floor upper bounds for a fixed number of bases} \label{supp:basis-ceilings}
We prove \pref{prop:basis-ceiling} and the End Matter refinements \eqsref{eq:weighted-ceiling}{eq:simple-basis-count}. We first bound the Fisher floor by evaluating the Fisher information at a vector of the most heavily weighted basis. We then use mean Fisher anisotropy to obtain a stronger bound in large dimensions. Both arguments apply to arbitrary choices of orthonormal bases and their selection probabilities.
\begin{proof}[Proof of \pref{prop:basis-ceiling}]
Write $q_\ell$ for the probability assigned to basis $\ell$, and let $q_{\max}:=\max_\ell q_\ell$. Choose $\ket\psi$ to be a vector of a basis with weight $q_{\max}$. At this state, that basis has exactly one outcome of nonzero probability, so its Fisher projector is zero. Every other basis has a Fisher projector of rank at most $d-1$. By \eref{eq:fisher-operator}, the trace of the normalized Fisher operator $G_\psi$ is therefore at most $(d-1)(1-q_{\max})$. Its smallest eigenvalue is at most its trace divided by $2d-2$, which gives \eref{eq:weighted-ceiling}. Since $q_{\max}\ge 1/L$, this proves the $1/L$ term in \eref{eq:finite-ceiling}. For upper bounds on the Fisher floor of more general finite POVMs, see Ref.~\cite{YangXiaoZhu2026}.

For $L\le2$, at most one basis contributes at this state. Thus $G_\psi$ has rank at most $d-1$ on the real tangent space of dimension $2d-2$. Its smallest eigenvalue is zero, so $\eta(\caM)=0$.

By \eref{eq:phase-paired-povm}, every Fisher eigenvalue lies in $[\eta(\caM),1-\eta(\caM)]$. Averaging its squared deviation from $1/2$ gives $\wp(\caM)\le[1/2-\eta(\caM)]^2$. The weighted extension of \pref{prop:mean-anisotropy}, proved in SM \sref{supp:average-mub}, gives, with $\varsigma:=\sum_{\ell=0}^{L-1}q_\ell^2$,
\begin{equation} \label{eq:weighted-floor-obstr}
4\wp(\caM)\ge1-(1-\varsigma)d\ln\frac{d}{d-1}.
\end{equation}
Together with $\wp\ge0$, these inequalities prove the sharper weight-dependent bound
\begin{equation} \label{eq:finite-settings}
\eta(\caM)\le\frac{1}{2}-\frac{1}{2}\sqrt{\max\left\{0,1-(1-\varsigma)d\ln\frac{d}{d-1}\right\}}.
\end{equation}
For a simpler finite-dimensional correction, observe that
\begin{equation} \label{eq:dimension-correction}
d\ln\frac{d}{d-1}-1=\int_0^1\frac{t}{d-t}\,\rmd t\le\frac{1}{2(d-1)}.
\end{equation}
Using $\varsigma\ge1/L$ in \eref{eq:weighted-floor-obstr} and applying \eref{eq:dimension-correction}, we obtain
\begin{equation} \label{eq:simple-anisotropy}
4\wp(\caM)\ge\frac{1}{L}\left[1-\frac{L-1}{2(d-1)}\right].
\end{equation}
Since $\sqrt{\max\{0,1-x\}}\lsp \ge1-x$ for $x\ge0$, this bound and $\wp\le(1/2-\eta)^2$ give the second term in \eref{eq:finite-ceiling}; the universal ceiling supplies the first. If $\eta\ge1/2-\varepsilon$, the universal ceiling gives $L\ge1/(2\varepsilon)$, while $4\wp\le4\varepsilon^2$ and \eref{eq:simple-anisotropy} give the other term in \eref{eq:simple-basis-count}.
 Finally $\varsigma\ge1/L$ and $d\ln[d/(d-1)]\rightarrow1$ prove \eref{eq:fixed-settings-ceiling}, even with probabilities that vary with $d$. Thus a floor $1/2-\varepsilon$ in arbitrarily large dimensions requires $L\ge1/(4\varepsilon^2)$.
\end{proof}

\subsection{Proof of \texorpdfstring{\pref{prop:frame-cert}}{Proposition 8}: a computable lower bound on the Fisher floor} \label{supp:deterministic-certs}

This subsection proves \pref{prop:frame-cert}. It also records the rank-two certificate as an additional observation. Use the equal mixture $\caM$ in \pref{prop:frame-cert}. Any verified bound $\Pi_{\caM}(O)\ge r\ge0$ for every $O\in\Xi_d$ certifies $\eta(\caM)\ge r^2$ by \eref{eq:rank-two-fisher}. Minimizing over this orbit remains a global optimization problem.

\begin{proof}[Proof of \pref{prop:frame-cert}]
A simpler sufficient test uses one matrix eigenvalue. Put $P_{\ell k}:=\ket{\varphi_{\ell k}}\bra{\varphi_{\ell k}}$ and define the self-adjoint frame operator
\begin{equation} \label{eq:frame-cert}
\sfS_{\caM}(O):=\frac{1}{L}\sum_{\ell,k}\tr(P_{\ell k}O)P_{\ell k},\quad
s_{\caM}:=\lambda_{\min}\left(\sfS_{\caM}\big|_{\tr O=0}\right),\quad \eta(\caM)\ge s_{\caM}.
\end{equation}
The restriction is to the real space of traceless Hermitian matrices with inner product $\langle O,Y\rangle=\tr(OY)$ and Hilbert--Schmidt norm $\|O\|_2=\sqrt{\tr\left(O^2\right)}$.
For an orthonormal pair $\ket\psi,\ket\nu$, put $O=\ket\psi\bra\nu+\ket\nu\bra\psi$ and $x_{\ell k}=\tr(P_{\ell k}O)$. Applying \eref{eq:single-basis-quadratic} in each basis gives
\begin{equation}
g_{\caM}(\psi,\nu)\ge\frac{1}{2L}\sum_{\ell,k}x_{\ell k}^2=\frac{1}{2}\langle O,\sfS_{\caM}(O)\rangle\ge s_{\caM},
\end{equation}
where the last inequality uses $\|O\|_2^2=2$. Taking the infimum proves the certificate. For each basis, the map $O\mapsto\sum_k\tr(P_{\ell k}O)P_{\ell k}$ projects orthogonally onto the operators diagonal in that basis because the $P_{\ell k}$ are Hilbert--Schmidt orthonormal. The diagonal space has dimension $d$. The zero-trace constraint removes one dimension, so the restricted projector has rank and trace $d-1$. Averaging the restricted projectors gives trace $d-1$ and rank at most $L(d-1)$. Thus $s_{\caM}\le1/(d+1)$ and $s_{\caM}=0$ if $L<d+1$. For equally weighted complete MUBs, the traceless diagonal subspaces are mutually orthogonal and span the traceless Hermitian space, as in the proof of \pref{prop:mub-floor}. Their projectors sum to the identity on that space, so $s_{\caM}=1/(d+1)=\eta(\caM)$.

\end{proof}

\section{Fisher floors of Haar-random bases} \label{supp:average-floors}
We now consider the random measurement $\caM_{d,L}$. The first two subsections bound its expected Fisher floor, using the mixing result in \sref{supp:mixing} for monotonicity. The last subsection proves a lower bound on the probability that its floor falls below a prescribed near-half target. All expectations and probabilities in this section concern the initial draw of the bases.

\subsection{Proof of \texorpdfstring{\pref{prop:mean-floor}}{Proposition 3}: expected Fisher floor} \label{supp:mean-floor-proof}
\begin{proof}[Proof of \pref{prop:mean-floor}]
Monotonicity follows from \pref{prop:mean-monotone}. To prove \eref{eq:mean-floor}, put $\Delta:=1/2-\eta(\caM_{d,L})\in[0,1/2]$ and $u:=\ln(\rme L)$. The bound is immediate when $L\le24u$, because its right-hand side is nonpositive. Otherwise, set
\begin{equation} \label{eq:mean-floor-parameters}
w=6u-1\ge5,\quad \varepsilon=\sqrt{\frac{w}{L}}\lsp <\frac{1}{2}.
\end{equation}
Since $\ln w\ge\ln5>7/6$,
\begin{equation}
12\ln\frac{\rme}{\varepsilon}
=6(1+u-\ln w)\le6u-1=w=L\varepsilon^2.
\end{equation}
Thus the setting-count condition in \thref{thm:near-half} holds at $\varepsilon$ and at every $x\in[\varepsilon,1/2)$. Applying that theorem and integrating the tail of $\Delta^2$ gives
\begin{equation} \label{eq:mean-floor-integrated}
\begin{aligned}
\E\Delta^2&\le\varepsilon^2+\int_{\varepsilon^2}^{1/4}\exp\left[-\frac{L(d+16)t}{24}\right]\,\rmd t\le\frac{w}{L}+\frac{24}{L(d+16)}\exp\left[-\frac{(d+16)w}{24}\right]\\
&\le\frac{w+(4/3)\rme^{-3w/4}}{L}\le\frac{w+1}{L}=\frac{6u}{L}.
\end{aligned}
\end{equation}
The last line uses $d+16\ge18$ and $w\ge5$. Cauchy--Schwarz now yields
\begin{equation}
\overline{\eta}_{\mathrm H}(d,L)
=\frac{1}{2}-\E\Delta
\ge\frac{1}{2}-\sqrt{\E\Delta^2}\lsp \ge\frac{1}{2}-\sqrt{\frac{6\ln(\rme L)}{L}},
\end{equation}
proving \eref{eq:mean-floor}.
\end{proof}

\subsection{Alternative bounds on the expected Fisher floor} \label{supp:moderate-floors}
\begin{proposition}[Expected floors at moderate basis counts] \label{prop:moderate-floors}
For integers $d\ge2$ and $L\ge1$, write $\overline{\eta}_{\mathrm H}(d,L):=\E\eta(\caM_{d,L})$ as in \pref{prop:mean-floor}. Then
\begin{align}
\overline{\eta}_{\mathrm H}(d,L)&\ge c_L\left(1-\rme^{-(d+18)/2}\right),\quad L\ge5, \label{eq:mean-floor-five}\\
\overline{\eta}_{\mathrm H}(d,L)&\ge\frac{1-\rme^{-L(d+10)/60}}{10},\quad L\ge71, \label{eq:mean-floor-tenth}\\
\overline{\eta}_{\mathrm H}(d,L)&\ge\frac{1}{4}-\frac{\rme^{-L(d+10)/60}}{10}-\frac{3\rme^{-L(d+10)/271}}{20},\quad L\ge271. \label{eq:mean-floor-quarter}
\end{align}
Here $c_L$, used only for $L\ge5$, is the constant in \eref{eq:five-basis-bound}.
\end{proposition}
\begin{proof}[Proof of \pref{prop:moderate-floors}]
Multiplying the guarantees in \thref{thm:five-bases} and the first part of \thref{thm:large-floors} by their respective floors proves \eqsref{eq:mean-floor-five}{eq:mean-floor-tenth}. For $L\ge271$, use the pointwise inequality
\begin{equation}
\eta\ge\frac{1}{10}\bbone_{\left\{\eta\ge\frac{1}{10}\right\}}+\frac{3}{20}\bbone_{\left\{\eta\ge\frac{1}{4}\right\}}.
\end{equation}
Taking expectations and applying both parts of \thref{thm:large-floors} proves \eref{eq:mean-floor-quarter}. These estimates can be more informative than \eref{eq:mean-floor} when $L$ is moderate.
\end{proof}

\subsection{Lower-tail probability of the Fisher floor of Haar bases} \label{supp:haar-converse}
We prove \pref{prop:haar-converse}. A lower-tail event at one fixed state--direction pair is enough to obstruct a uniform floor. The Beta distribution gives its second and fourth moments, and Cantelli's inequality converts them into the required probability bound.
\begin{proof}[Proof of \pref{prop:haar-converse}]
Fix orthonormal $\ket\psi,\ket\nu\in\caH$ and put $\xi_\ell:=g_{\caB_\ell}(\psi,\nu)-1/2$, $\ell=0,\ldots,L-1$. Independence of the bases makes these variables independent. The symmetric Beta distribution in \lref{lem:beta-laplace} gives continuous distributions with mean zero and moments
\begin{equation} \label{eq:haar-moments}
\E \xi_\ell^2=\frac{1}{4d},\quad \E \xi_\ell^4=\frac{3}{16d(d+2)}.
\end{equation}
For $\xi:=\sum_{\ell=0}^{L-1}\xi_\ell$, independence and zero means give
\begin{equation} \label{eq:haar-sum-moments}
\E \xi^2=\frac{L}{4d},\quad \E \xi^4=\left[3-\frac{6}{L(d+2)}\right]\left(\E \xi^2\right)^2.
\end{equation}
For this fixed state, all basis overlaps are nonzero almost surely. The phase-pair equality in \eref{eq:phase-paired-resp} therefore gives responses $1/2+\xi/L$ and $1/2-\xi/L$, so
\begin{equation}
\eta(\caM_{d,L})\le\frac{1}{2}-\frac{|\xi|}{L}.
\end{equation}
Apply Cantelli's one-sided variance inequality to $\xi^2$, whose mean and variance follow from \eref{eq:haar-sum-moments}. Since $L^2\varepsilon^2<\E\xi^2$, it gives
\begin{equation}
\Pr\left\{\xi^2>L^2\varepsilon^2\right\}
\ge\frac{(\E\xi^2-L^2\varepsilon^2)^2}{\E\xi^4-(\E\xi^2)^2+(\E\xi^2-L^2\varepsilon^2)^2}.
\end{equation}
Together with the preceding floor bound, this gives the sharper estimate
\begin{equation} \label{eq:haar-cantelli}
\Pr\left\{\eta(\caM_{d,L})<\frac{1}{2}-\varepsilon\right\}
\ge\frac{\left(1-4Ld\varepsilon^2\right)^2}{2-\frac{6}{L(d+2)}+\left(1-4Ld\varepsilon^2\right)^2}.
\end{equation}
For $0<\varepsilon<1/(2\sqrt{Ld}\lsp)$, we have $0<1-4Ld\varepsilon^2<1$, so the denominator is at most $3$. This proves the simpler End Matter bound \eref{eq:anti-concentr}. To obtain the necessary count, use the sharper estimate \eref{eq:haar-cantelli}. If $Ld\varepsilon^2\le7/40$, then \eref{eq:haar-cantelli} applies and $\left(1-4Ld\varepsilon^2\right)^2\ge9/100$. The right-hand side is increasing in this squared term and exceeds $9/209>1/24$. Taking $d=2$ shows that $L\le7/(80\varepsilon^2)$ precludes the stated success probability.
\end{proof}

\section{Data and numerical methods} \label{supp:figure-certs}
The upper panel of \fref{fig:anisotropy-floor} shows the value for complete MUBs, for which \eref{eq:average-deviation} holds with equality, and the Haar expectation $1/[4(d+1)]$ from \eref{eq:haar-average-deviation}. In the lower panel, the floor for complete MUBs with equal weights is $1/(d+1)$ by \pref{prop:mub-floor}. The horizontal line shows the lower bound $1/10$ from \thref{thm:large-floors} for $L=71$ Haar bases. This bound holds with probability at least $99.99\%$ in each dimension $d\ge2$. For the Haar points with $L=d+1$ in the lower panel, we evaluate the applicable bounds from Theorems~\ref{thm:five-bases}--\ref{thm:near-half}. At each dimension, we plot the largest of these lower bounds whose success probability over the draw of the Haar bases is guaranteed to be at least $99.99\%$. This gives positive bounds at 51 of the 53 dimensions shown. The Haar points with $L=d+1$ are omitted at $d=2$ and $d=3$. For \thref{thm:near-half}, we use bisection to find the value of $\varepsilon$ for which the sufficient number of bases equals $L$. We round the resulting lower bound on the floor down and check the conditions of the theorem using outward-rounded interval arithmetic. We round the reported upper bounds on the failure probability up.

\Fref{fig:accuracy-count} shows the number of bases sufficient by \thref{thm:near-half} and the number necessary for arbitrarily large dimensions by \pref{prop:basis-ceiling}. Both curves are computed from the unrounded formulas for $3\le 1/\varepsilon\le53$. For the numerical points, we fix $d=32$ and use 100 independent measurements for each of 21 values of $L$. Each measurement gives equal weight to $L$ independent Haar bases. We use 800 starting states per measurement. Of these, 400 are the states with the smallest values of $\lambda_{\min}(G_\psi)$ in a pool of 1,000 Haar-random pure states, and the other 400 are independently drawn Haar-random pure states. We evaluate the candidate states using the original Fisher matrix and keep the smallest eigenvalue found as a numerical upper estimate of the floor of that measurement. We average the 100 estimates to obtain $\bar{\eta}_L$ and plot the point $(1/(1/2-\bar{\eta}_L),L)$.

\emph{Five Haar bases.} In each dimension from 2 to 100, we draw 100 independent measurements. Each measurement gives equal weight to five independent Haar bases. We use 800 starting states per measurement. The smallest value of $\lambda_{\min}(G_\psi)$ found in the search is taken as a numerical upper estimate of the Fisher floor of that measurement. We then average the 100 estimates in each dimension.

\makeatletter\let\auto@bib@innerbib\@empty\let\auto@bib\@empty\makeatother
\end{document}